\documentclass[superscriptaddress,12pt,tightenlines,aps,nofootinbib,pre]{revtex4-1}
\usepackage{color,epsfig,graphics,amssymb,amsmath,subeqnarray,graphicx,amsthm,subfigure,mathrsfs,bm,color,xcolor,soul}
\usepackage{natbib}

\def\createbib #1{\bibliographystyle{jsk}
\bibliography{#1}}

\usepackage{natbib}
\usepackage{graphicx}
\usepackage{amsmath}
\usepackage{amssymb}
\usepackage{natbib}
\usepackage{soul}
\usepackage{color}
\def\bea{\begin{equation}}
\def\eea{\end{equation}}
\def\Real{\mbox{Re}}

\newcommand{\diff}[2]{\frac{\mbox{d} #1}{\mbox{d} #2}}
\newcommand{\pdiffn}[3]{\frac{\partial^{#3} #1}{\partial #2^{#3}}}
\newcommand{\pdiff}[2]{\frac{\partial #1}{\partial #2}}

\begin{document}

\title{Stability and Bifurcation of Dynamic Contact Lines in Two Dimensions}
\author{J. S. Keeler}
\affiliation{Mathematics Institute, University of Warwick,
Coventry, CV4 7AL, UK}
\email{jack.keeler@warwick.ac.uk}
\author{D. A. Lockerby}
\affiliation{School of Engineering, University of Warwick,
  Coventry, CV4 7AL, UK}
\author{S. Kumar}
\affiliation{Department of Chemical Engineering and Materials Science, University of Minnesota, Minneapolis, MN 55455, USA}
\author{J. E. Sprittles}
\affiliation{Mathematics Institute, University of Warwick,
Coventry, CV4 7AL, UK}

\vspace{1cm}






\begin{abstract}
The moving-contact line between a fluid, liquid and a solid is a ubiquitous phenomenon, and determining the maximum speed at which a liquid can wet/dewet a solid
is a practically important problem. Using continuum models, previous studies have shown that the maximum speed of wetting/dewetting can be found by calculating steady solutions of the governing equations and locating the critical capillary number, $Ca_{\mathrm{crit}}$, above which no steady-state solution can be found. Below $Ca_{\mathrm{crit}}$, both stable and unstable steady-state solutions exist and if some appropriate measure of these solutions is plotted against $Ca$, a fold bifurcation appears where the stable and unstable branches meet. Interestingly, the significance of this bifurcation structure to the transient dynamics has yet to be explored. This article develops a computational model and uses ideas from dynamical systems theory to show the profound importance of the unstable solutions on the transient behaviour. By perturbing the stable state by the eigenmodes calculated from a linear stability analysis it is shown that the unstable branch is responsible for the eventual dynamical outcomes and that the system can become unstable when $Ca<Ca_{\mathrm{crit}}$ due to finite amplitude perturbations. Furthermore, when $Ca>Ca_{\mathrm{crit}}$, we will show that the trajectories in phase space closely follow the unstable branch.
\end{abstract}
\maketitle
\section{Introduction}

\begin{figure}
\centering
  \includegraphics[scale=0.7]{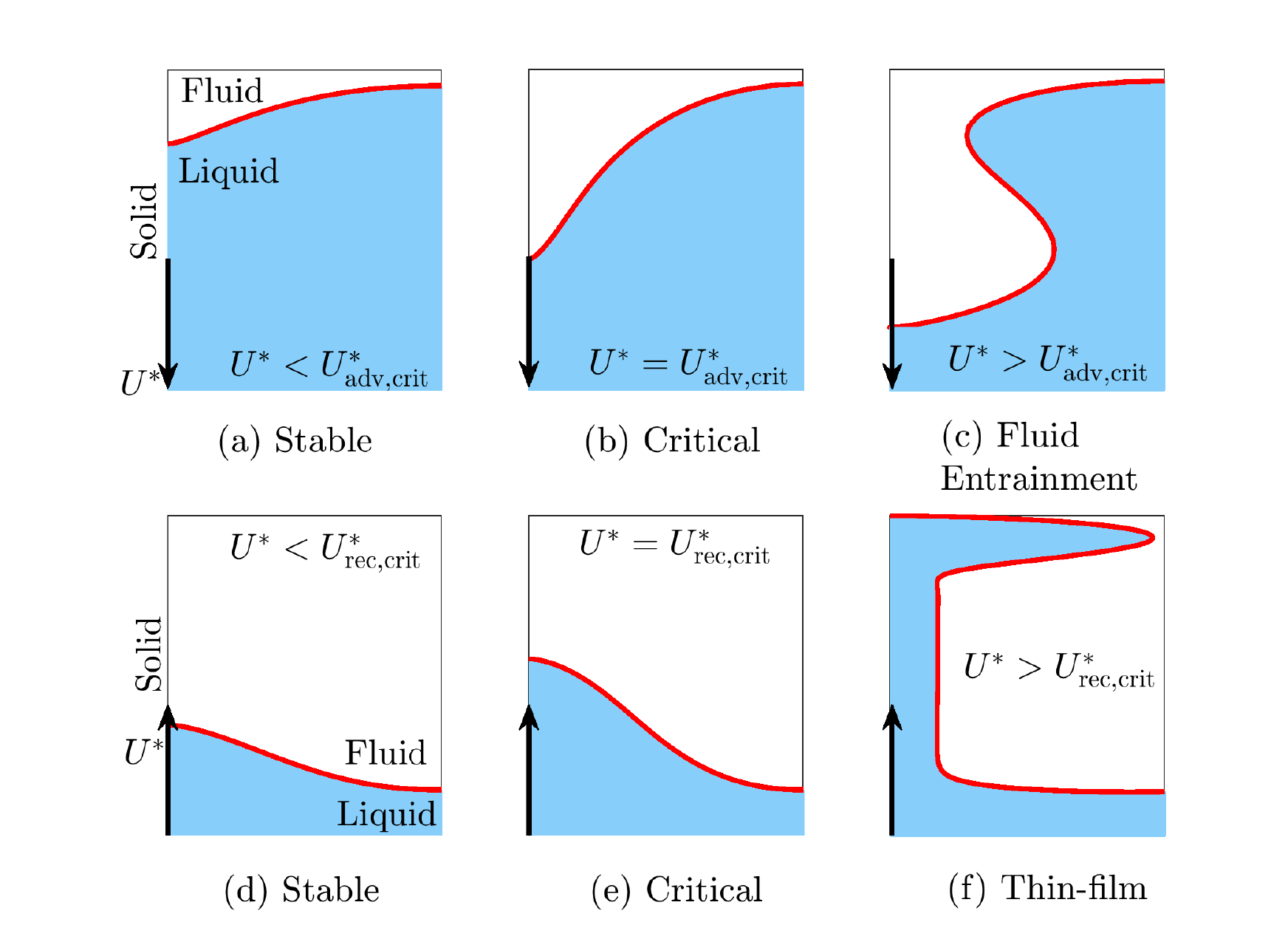}
	\caption{The moving contact line problem in a channel geometry in a frame of reference that moves with the liquid. Panels (a)-(c) describe the advancing contact line problem whilst (d)-(f) describe the receding contact line problem. In both cases as the speed of the substrate, $U^*$, increases the system is first stable (panels (a) and (d)) before the system becomes unstable and air entrainment (panel (c)) or a thin-film formation (panel (f)) occurs.}
  \label{fig:system_sketch}
\end{figure}

Understanding the shape and evolution of the interface between a fluid, liquid and a solid substrate is a classic problem in fluid mechanics and yet a remarkable number of open questions still remain \citep{afkhami2020challenges,velarde2011discussion}. There are two fundamental cases: an advancing contact line, where a liquid phase advances and `wets' the solid, see figure~\ref{fig:system_sketch}(a)-(c), and a receding contact line, where a liquid phase recedes and `dewets' the solid, see figure~\ref{fig:system_sketch}(d)-(f). Both experimental and theoretical studies \citep[][]{bonn2009wetting,snoeijer2013moving} have shown that there is a critical contact line speed relative to the solid, beyond which stability is lost and the system ceases to return to a steady state. In the case of an advancing contact line (see figure~\ref{fig:system_sketch}(c)) this instability is characterised by fluid entrainment (which in many practical cases is air entrainment) and in a receding contact line (see figure~\ref{fig:system_sketch}(f)) a thin liquid film is deposited on the solid. The principle aim of this article is to provide insight into this instability and understand the dynamics of the system near the critical speed.

\begin{figure}
\centering
  \includegraphics[scale=0.5]{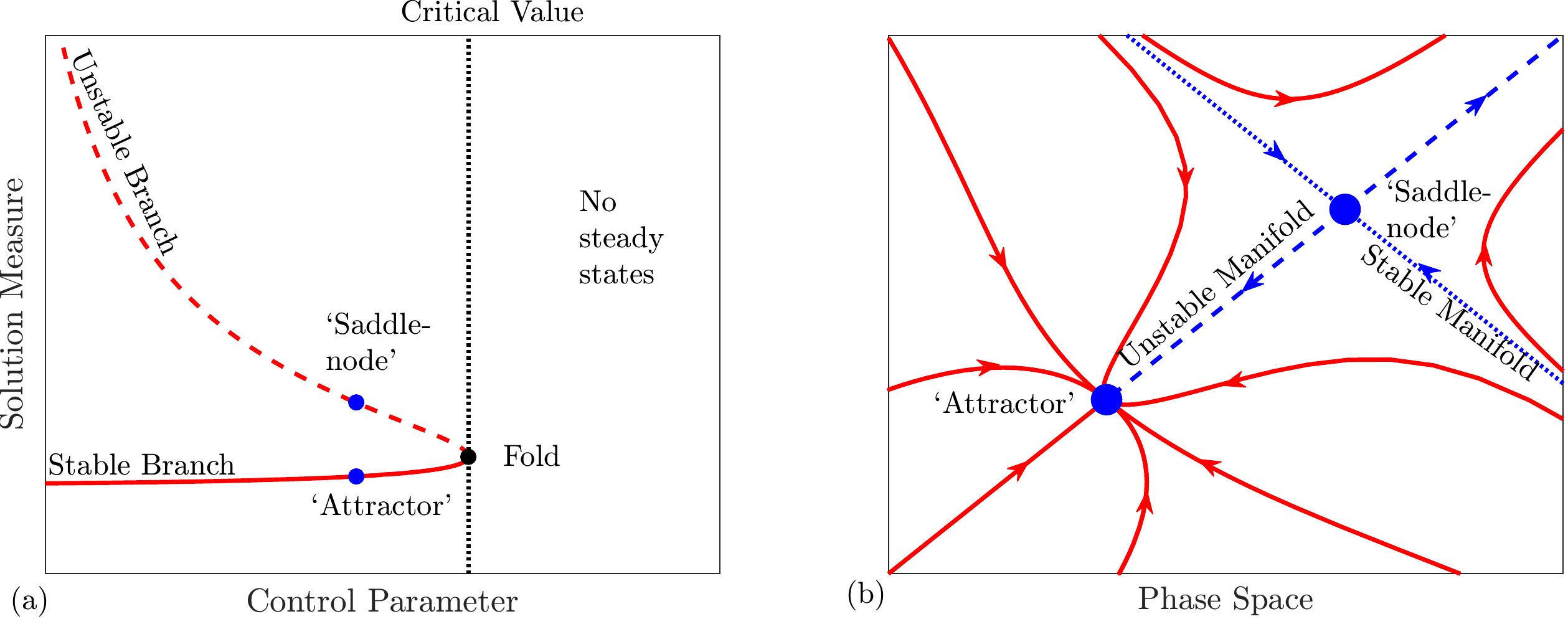}
	\caption{(a) A sketch of a typical fold bifurcation structure. A solution measure is plotted against a control parameter to form a `solution curve'. At a critical value, two branches -- one stable (solid line) and one unstable (dashed line ---) meet. The location of their intersection is known as a fold bifurcation. Beyond the critical value there are no (known) steady states. In our specific problem the control parameter is $Ca$ and the solution measure is either the interface length or meniscus rise. (b) A generic two-dimensional phase plane for a parameter value less than the critical value with a stable state (an `attractor' on the stable branch, see (a)) and an weakly unstable state (a `saddle-node' on the unstable branch, see (a)). The unstable/stable manifolds of the unstable state are dashed/dotted respectively.}
  \label{fig:bifurcation_sketch}
\end{figure}

The critical speed where the instability occurs is associated with a fold bifurcation in the steady solution structure, which divides the steady solutions between a stable branch and an unstable branch (figure~\ref{fig:bifurcation_sketch}(a) and see \cite{kuznetsov2013elements} for a detailed mathematical description). For parameter values `beyond the fold' there are no (known) two-dimensional steady states and the system must becomes transient and/or three-dimensional. In our system the appropriate non-dimensional parameter associated with the speed of the solid is the capillary number, $Ca$ (see next section for a precise definition). Whilst analysis of the unstable branch of solutions (which exists for parameter values `below the fold') can reveal important information about transient behaviour, the focus of theoretical studies has been mainly to calculate and characterise only the stable steady solutions immediately up to the critical speed \citep[see, for example][]{eggers2005existence,snoeijer2012theory,vandre2013mechanism,sprittles2015air}. However, \cite{snoeijer2012theory} hypothesised that the set of unstable solutions represents what they termed `effective dynamics', i.e. the system's time-dependent trajectory for $Ca>Ca_{\mathrm{crit}}$ closely matches the unstable branch when plotted using appropriate measures. If so, the unstable branch is not just an insignificant consequence of the fold bifurcation but provides unique insight into the system dynamics. The influence and importance of unstable states in fluid dynamics systems has been investigated in many different contexts, including shear flow \citep{eckhardt2008dynamical}, droplets \citep{gallino2018edge}, finite air bubbles \citep{keeler2019invariant,gaillard2020life} and a slide-coating flow \citep{christodoulou1988finding}. Indeed, as shown in figure~\ref{fig:bifurcation_sketch}(b), where the phase-plane is sketched for a generic system with an stable (`attractor') and weakly unstable (`saddle-node') state, the unstable state can act as a separator of dynamical outcomes; the stable manifold is a dividing `line' and the unstable manifold connects to the stable state. In this article we adapt these dynamical systems ideas to the moving-contact-line problem to reveal the role of the unstable solutions. We calculate the bifurcation structure and stability properties of the steady solutions and relate these to time-dependent calculations in the sub-critical ($Ca<Ca_{\mathrm{crit}}$) and super-critical ($Ca>Ca_{\mathrm{crit}}$) regimes.

We now provide some important background on moving contact lines. It is well known that the classical `moving contact line paradox', as described in \cite{huh1971hydrodynamic} can be alleviated if there is slip near the contact point. If this slip occurs in an inner region, as considered by \cite{voinov1976} and \cite{cox1985part1}, then bending of the interface occurs in an intermediate region where viscous effects can cause the liquid-fluid interface to curve sharply. In this formulation, it is often assumed that the intermediate region connects to an outer region where the interface retains its static meniscus shape. The possible asymptotic matching of these regions has critical consequences and provides insight into the bifurcation structure of the steady solution space. In a series of remarkable articles, it was shown, by solving a lubrication model for a liquid-vacuum system, how the curvature of the inner and outer regions can be asymptotically matched. For the advancing contact line this can be achieved for all values of $Ca$, but for the receding contact line, the matching fails when $Ca$ is past some critical threshold, interpreted as $Ca_{\mathrm{crit}}$ \citep{eggers2004towards,eggers2004forced,eggers2005existence}. The bifurcation structure of the stable and unstable branches of the receding contact line was then fully described using matched asymptotics and bifurcation theory by \cite{snoeijer2012theory} for $Ca\ll 1$, and $Ca_{\mathrm{crit}}$ was determined to occur at a fold bifurcation. 

The aforementioned lubrication analysis has been extended to general liquid-fluid systems, where the viscosity of the fluid phase is considered non-zero \citep{chan2020cox,kamal2019dynamic,chan2013hydrodynamics} and also for the full Navier-Stokes equations \citep{vandre2012delay,vandre2013mechanism,vandrethesis}. A key result from these studies is that for the advancing contact line the presence of viscosity fundamentally alters the bifurcation structure and a fold bifurcation appears at a finite $Ca$. \cite{vandre2013mechanism} showed that the fold bifurcation in the advancing contact line problem occurs when the horizontal air pressure gradient matches the strength of capillary-stress gradient near the contact point. It was also demonstrated that using the lubrication model, in both phases, poorly predicts $Ca_{\mathrm{crit}}$ when compared to the full Navier-Stokes equations for the advancing contact line \citep{vandre2012delay,vandre2013mechanism,vandrethesis}. Other physical effects such as Maragoni flows, inertia and gravity, and shear thinning/thickening were also found to preserve the fold bifurcation \citep{vandre2013mechanism,liu2019predictions,liu2016assist,liu2016surfactant,liu2017mechanism,charitatos2020dynamic}. 

In physical terms, in the advancing case, the critical behaviour indicates the threshold at which fluid entrainment occurs where, experimentally, a three-dimensional saw-tooth pattern emerges as observed in a variety of different flow configurations e.g. liquid films \citep{reysatt2006burst}, drop impact \citep{thoroddsen2012micro,pack2018contact} and plate penetration in a liquid bath \citep{he2019characteristic}. In the receding case, however, the fold bifurcation marks the onset of thin-film deposition \citep{snoeijer2006avoid,snoeijer2008thick}. Interestingly, despite the 3D structures of air entrainment \citep{he2019characteristic,he2020long}, 2D models appear to accurately predict the transition point, an observation which is yet to be understood \citep[see, for example][]{liu2019predictions,vandre2012delay,sprittles2017kinetic}. Transversal three-dimensional perturbations have been considered for the receding contact line \citep{snoeijer2007part1} and the advancing contact line \citep{vandrethesis}, both using a lubrication model, but a stability analysis using the full hydrodynamics equations has not yet been conducted.

In this article we consider the bifurcation structure devoid of any limitations from the lubrication approximation, allowing us to naturally consider both advancing and receding cases simultaneously. Our analysis of two-phase contact-line stability will focus on steady-state solutions using a hybrid model; the liquid phase is modelled using the Navier-Stokes equations and the fluid phase is accurately modelled using a lubrication approximation \citep[see][]{liu2019predictions,liu2016assist,liu2016surfactant,liu2017mechanism,stay2003coupled,sprittles2017kinetic}. 

The structure of the article is as follows. In \S~\ref{sec:model} we describe the hydrodynamic equations that describe the system. In \S~\ref{sec:stability} we calculate the steady solution curves to determine the critical parameters associated with the loss of stability of the system. In addition, we perform a numerical linear stability analysis that reveals the significance of the unstable branch to the transient dynamics of the system. By treating the governing equations as a dynamical system we form a generalised eigenproblem that can be solved numerically to determine and quantify the stability of the solution branch. Next, in \S~\ref{sec:transient_dynamics}, by solving a time-dependent initial value problem (IVP) numerically we are able to demonstrate that far from having a passive role, the unstable branch represents, in the language of dynamical systems, the `basin boundary of attraction' of the stable state. Furthermore, by examining the phase-plane of the solution trajectory, we discover that the subsequent unsteady time-evolution is intrinsically linked to the unstable branch and are able to confirm the prediction of \cite{snoeijer2012theory} that the solution moves quasi-statically along the unstable branch. Viewing the trajectories through the lens of the phase plane will also allow us to understand if, and how, the system becomes unstable when $Ca<Ca_{\mathrm{crit}}$ and also provide criteria that could potentially enable suppression of this instability. Finally, in \S~\ref{sec:discuss}, we discuss the implications of these results and some possible future research. 

\section{Governing Equations}\label{sec:model}

\begin{figure}
\centering
  \includegraphics[scale=0.4]{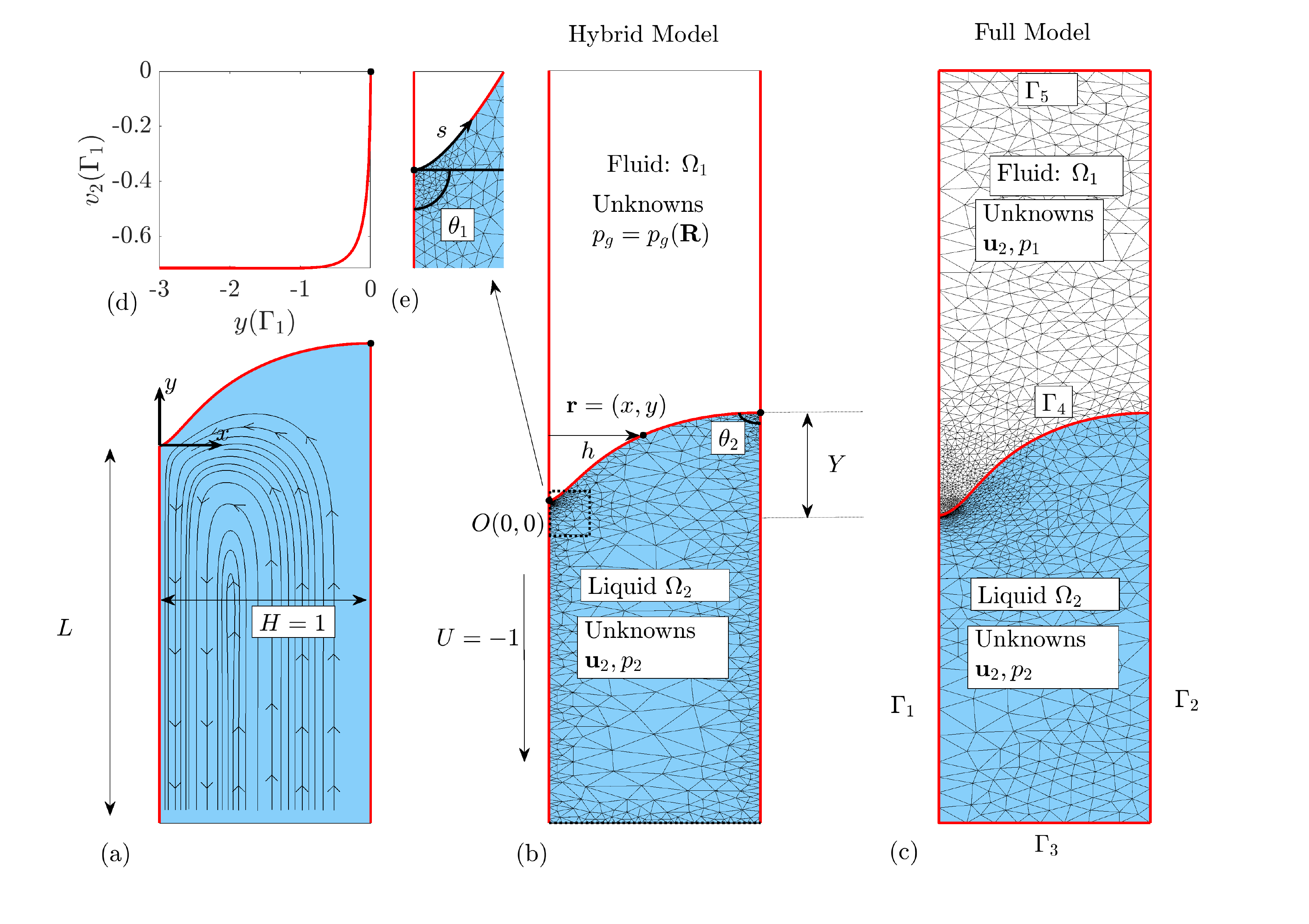}
  \caption{The computational domain for the hybrid and full models for an advancing contact line. The boundaries are denoted by $\Gamma_i$ (labelled in panel (c)) with the origin centred on the contact line. (a) A typical streamline pattern for a steady solution in the hybrid model. (b) The computational domain for the hybrid model.  In this model we solve for the velocity and pressure in the liquid domain, but only solve for the pressure of the fluid on the interface boundary, $\Gamma_4$. (c) The computational domain for the full model where the velocity and pressure fields are also solved in the fluid domain. (d) The vertical component of velocity along the moving wall, i.e. $\Gamma_1$. (e) The enlargement near the contact line shows the mesh refinement required to ensure that the flow-field is sufficiently resolved.}
  \label{fig:domain}
\end{figure}

We now discuss the hydrodynamic model and the assumptions that allow us to derive an accurate simplified hybrid model that is used in the calculations thereafter. The following discussion applies to both the advancing and receding contact lines, although the demonstrative figures only show the advancing contact line.

\subsection{Full Hydrodynamic Model}

Motivated by the system used in \cite{vandre2012delay}, we consider two-dimensional flow between two parallel plates, as shown in figure~\ref{fig:system_sketch}. In the following discussion we denote dimensional quantities using a $*$ superscript. Two fluids of viscosity $\mu_{1,2}^*$, and density $\rho_{1,2}^*$, fill the channel bounded by two rigid plates which are separated by a fixed height $H^*$, subscripts with 1 indicate the upper fluid (the fluid phase) and 2 indicates the lower fluid (the liquid phase). In our system the left plate moves with constant speed in the $y-$direction $U^*$ and the right plate is stationary. For a receding contact line $U^*>0$ and an advancing contact line $U^*<0$. The left wall is moving with a speed $U^*$ and the right wall is stationary. The coordinate system is centred on the contact point between the two fluids and the left (moving wall), see panel (a) of figure~\ref{fig:domain}. The fluid flow of each phase is governed by the two-dimensional Navier-Stokes equations. All speeds, lengths, pressures and times are scaled by $U^*$, $H^*$, $\mu_2^*U^*/H^*$ and $H^*/U^*$ respectively. Finally the viscosity ratio, denoted $\chi$, is defined with respect to the liquid phases, i.e. $\chi = \mu_{1}^*/\mu_{2}^*$.

As in previous studies \citep{vandre2013mechanism,vandre2012delay,liu2019predictions,liu2016assist,liu2016surfactant,liu2017mechanism,sprittles2013finite,sprittles2011viscous1,sprittles2011viscous2} we apply the Stokes-flow approximation so that the Reynolds number, $Re = U^*H^*\rho_2^*/\mu_2^*$, is negligible and assumed zero; results in \cite{vandre2013mechanism} show $Re$ can have an influence at sufficiently high values but it does not qualitatively alter their conclusions. We assume that gravitational effects are negligible throughout. The non-dimensional computational domain is shown in figure~\ref{fig:domain}(c). The line corresponding to the left plate is denoted $\Gamma_1$, the right plate $\Gamma_2$, the bottom boundary $\Gamma_3$, the free-surface $\Gamma_4$ and the top boundary $\Gamma_5$. The fluid and liquid domains are denoted by $\Omega_1$ and $\Omega_2$ respectively. The Stokes-flow equations, for the fluid velocity, $\textbf{u}_i = (u_i,v_i)$, and pressure, $p_i$, in each phase can be written as
\begin{align}
\chi \nabla^2\textbf{u}_1 &=\: \nabla p_1,\qquad\nabla\cdot\textbf{u}_1 = 0,\qquad \textbf{x}\in \Omega_1,\qquad\mbox{(Fluid)}\label{fluid_phase}\\ 
\nabla^2\textbf{u}_2 &=\: \nabla p_2,\qquad\nabla\cdot\textbf{u}_2 = 0,\qquad \textbf{x}\in \Omega_2.\qquad\mbox{(Liquid)}\label{liquid_phase}
\end{align}

On the left (moving) and right (stationary) walls, $\Gamma_1$ and $\Gamma_2$ respectively, we implement a Navier-slip condition written as
\begin{align}
  \lambda ( \boldsymbol{\tau}_i \cdot \textbf{n})\cdot\textbf{t} &=\: (\textbf{u}_{i} - \textbf{U})\cdot \textbf{t},\qquad\textbf{x}\in\Gamma_{1},\qquad i = 1,2,\\
\lambda ( \boldsymbol{\tau}_i \cdot \textbf{n})\cdot\textbf{t} &=\: \textbf{u}_{i} \cdot \textbf{t},\qquad\textbf{x}\in\Gamma_{2},\qquad i = 1,2,
\label{navier_slip}
\end{align}
where $\textbf{n}$ and $\textbf{t}$ are the vectors normal and tangential to each wall, $\textbf{U} = (0,U)$ is the non-dimensional speed of the wall and $\lambda$ is the non-dimensional slip length which, for simplicity, we assume to be the same in each phase (see, \cite{sprittles2017kinetic} for potential extensions). We choose to implement a Navier-slip condition on the stationary wall for consistency with the Navier-slip condition on the moving wall, although we could fix $\textbf{u}_i = \textbf{0}$ on $\Gamma_2$ and get similar results (see, for example \cite{vandre2013mechanism,liu2017mechanism}). The stress tensor in each phase $\boldsymbol{\tau}_i$ is defined as
\bea
\boldsymbol{\tau}_i = -p_i\textbf{I} + \delta_i\left(\nabla \textbf{u}_i + (\nabla\textbf{u}_i)^T\right),
\eea
where $\textbf{I}$ is the identity matrix and $\delta_1 = \chi, \delta_2 = 1$. On the interface between the two fluids we assume a constant surface tension, $\gamma^*$, so that the dynamic boundary condition can be written as
\bea
\boldsymbol{\tau}_2\cdot\textbf{n} - \boldsymbol{\tau}_1\cdot\textbf{n} = \frac{1}{Ca}\kappa \textbf{n},\qquad\textbf{x}\in\Gamma_{4},
\label{dynamic_condition}
\eea
$\textbf{n}$ is the normal of the interface pointing towards the fluid phase, $\kappa = \nabla\cdot\textbf{n}$ is the curvature of the interface and $Ca = \mu_2^*|U|/\gamma^*$ is the capillary number. We denote the unknown position of the interface as $\textbf{r} = (x_s,y_s)$, see figure~\ref{fig:domain}(b), so that the kinematic condition on the interface can be written as
\bea
\pdiff{\textbf{r}}{t}\cdot\textbf{n} = \textbf{u}\cdot\textbf{n}, \qquad\textbf{x}\in\Gamma_{4}.
\label{kinematic_condition}
\eea

In addition, we have to specify the angle the interface makes on the lower and upper walls. These angles can be allowed to vary with the capillary number, slip-length or other quantities but we choose the simplest approach and choose constant values, i.e.
\begin{align}
  \theta &=\: \theta_1,\,\mbox{on}\:\Gamma_1\cap\Gamma_4,\\
  \theta &=\: \theta_2,\,\mbox{on}\:\Gamma_2\cap\Gamma_4.
\label{angle_condition}
\end{align}
It is straightforward to replace the conditions in \eqref{angle_condition} with an equation involving $Ca$ and other quantities, but this is not the focus of the article.

Finally, we implement fully-developed flow conditions on the inflow and outflow boundaries,
\bea
\textbf{u}_i\cdot\textbf{t} = 0,\,\textbf{x}\in\Gamma_{3}\cup\Gamma_{5},
\label{fully_developed_flow_condition}
\eea
alongside a pressure drop across the domain so that,
\begin{align}
  p_1 &=\: 0,\,\textbf{x}\in\Gamma_{5},\\
  p_2 &=\: p_{\mathrm{out}},\,\textbf{x}\in\Gamma_{3}.
\label{pressure_drop_equation}
\end{align}

The full hydrodynamic system is defined in \eqref{fluid_phase}-\eqref{pressure_drop_equation} with the following high-dimensional state vector (denoted $\textbf{w}$) of unknowns;
\bea
\textbf{w} = [\textbf{u}_1,\textbf{u}_2,p_1,p_2,\textbf{r}]^T.
\label{full_unknowns}
\eea
It is worth noting that we model the effect of varying the speed of the wall by varying $Ca$ and that the non-dimensional slip-length, $\lambda$, can be varied to investigate changes in physical channel width. Finally as we are interested in the contact line of the left plate, we set $\theta_2 = \pi/2$ on the right plate, in all simulations, for simplicity. Therefore we have a set of control parameters
\bea
Ca,\lambda,\chi,\theta_1,p_{\mathrm{out}},
\label{full_parameters}
\eea
that need to be specified in order to solve \eqref{fluid_phase}-\eqref{pressure_drop_equation}. 

\subsection{Hybrid Model}

The computational cost of the full model can be drastically reduced by solving the thin-film equations where they are valid \citep{sbragaglia2008wetting,jacqmin2004onset,oron1997long}, leading to a hybrid model \citep[see][]{liu2019predictions,liu2016assist,liu2016surfactant,liu2017mechanism,stay2003coupled} which approximately halves the complexity of the problem, as unknowns in the fluid phase are only computed on the interface. The difference of our approach from previous implementations is that our hybrid model takes into account time-dependence so that stability can be probed and IVP calculations can be performed. The key assumption is that typical horizontal length scales are small compared to typical vertical length scales so that the horizontal component of velocity in the fluid phase is small, i.e. $u_1\ll 1$. The full derivation is discussed in Appendix~\ref{app:hybrid} and the computational domain is shown in figure~\ref{fig:domain}(b).

The effect of this reduction in the fluid phase is to replace a full two-dimensional description, given in \eqref{fluid_phase}, by a one-dimensional equation for the fluid pressure, $p_1$, on the interface only. This equation can be stated as
\bea
\pdiff{h}{t} + \frac{1}{\chi}\pdiff{Q_1}{s},\qquad Q_1 = \frac{1}{6}\pdiff{p_1}{s}h^3 + \frac{1}{2}Ah^2 + Bh = 0,
\label{hybrid_equation}
\eea
where $h$ is the horizontal distance, from the left hand solid, to the interface (see figure~\ref{fig:domain}), $Q_1$ is the flux, and the constants $A$ and $B$ are functions of $\chi,\lambda$ and $\textbf{u}_2$ and are given in Appendix~\ref{app:hybrid}. The fluid phase is coupled to the liquid phase through the applied traction given in \eqref{coupling}. We now have a system of PDE described by equations \eqref{liquid_phase}-\eqref{pressure_drop_equation} and \eqref{hybrid_equation} with the high-dimensional state vector of unknowns:
\bea
\textbf{w} = [\textbf{u}_2,p_2,p_1(\textbf{r}),\textbf{r}]^T.
\label{hybrid_unknowns}
\eea
We validate the hybrid model by comparing to the full hydrodynamic model in Appendix~\ref{app:hybrid}. Finally we note that this approach is strictly only valid for the advancing contact line problem but, as shall be shown later, the receding contact-line problem is effectively a one-phase problem, (c.f. figure~\ref{fig:ca_crit_evolution}) and implementing the hybrid model for a receding contact line does not significantly change the value of $Ca_{\mathrm{crit}}$ (Appendix~\ref{app:hybrid}).

\subsubsection{System Parameters and Integral Measures}
We now describe additional system parameters and measures that will be useful in computing and describing the steady and time-dependent solutions. The pressure at the outflow boundary, $p_{\mathrm{out}}$, is determined by an integral volume constraint {acting on the liquid phase}, i.e.
\bea
\int_{{\Omega_2}} \mbox{d}V = V,
\label{volume_constraint}
\eea
where $V$ is the volume per unit length of the domain  (corresponding to the area of the computational domain). In our numerical calculations the position of the contact points on the moving and stationary plates are both allowed to move so that \eqref{volume_constraint} can be satisfied. For ease of presentation, we post-process and rescale the solution so that the origin is always at the contact point of the moving plate. In the calculations that follow we choose $V=5$ which, after careful experimentation, is large enough for fully-developed flow to occur near the outflow boundary, $\Gamma_1$. We find that the solutions are independent of values of $V\geq5$ that we choose. 

For a fixed {set of parameter values, as defined in \eqref{full_parameters},} we calculate steady solution curves by setting the time derivatives in the governing equations to zero and {then solving} the resulting steady system. As we vary $Ca$, and then subsequently calculate a solution, a solution curve will be traced and a fold bifurcation will occur at the critical value of the capillary number, denoted $Ca_{\mathrm{crit}}$. Whilst it is possible to trace a solution branch around the fold numerically by a pseudo-arclength continuation method \citep[see, for example][]{doedel}, we implement an alternative, bespoke approach. We expect the interface length, $L$, to increase monotonically as the curve is traced out around the fold and therefore is a suitable candidate for a continuation parameter that allows us to calculate solutions smoothly around the fold. To achieve this we let $Ca$ become an unknown parameter that is determined by setting the total length of the interface, i.e. 
\bea
\int_{\Gamma_{4}}\mbox{d}s = L.
\label{arclength_constraint}
\eea
This approach enables us to trace solution curves around the fold by incrementally increasing $L$ and solving the system of equations with $Ca$ determined by \eqref{arclength_constraint}. We also emphasise that \eqref{arclength_constraint} can only be implemented in steady calculations as a means of tracing the solution curve whereas the volume constraint in \eqref{volume_constraint} is applied in both steady and time-dependent calculations.

Finally, when describing the steady solutions and time-dependent solutions we use the meniscus rise (more specifically, the vertical distance between the two contact lines) defined as
\bea
Y(t) = |y(s=L) - y(s=0)|
\label{delta}
\eea
as a convenient solution measure (as previously considered, for example, in \cite{kamal2019dynamic}; see figure~\ref{fig:domain} (in this paper) for reference). 

\subsection{Numerical Method}\label{sec:num_method}
The governing equations are solved using the finite-element method from within the open-source \texttt{oomph-lib} object-orientated multi-physics library \citep{heil2006oomph}. The structure and implementation of our equations follows that of \cite{sprittles2013finite}. Following multiplication of the equations by a test function, $\psi$, and then an integration over the domain, the boundary integrals that result from integration by parts require the traction to be specified on each of the boundaries. The dynamic condition, \eqref{dynamic_condition}, and the Navier-slip condition, \eqref{navier_slip}, therefore can be implemented as a natural condition by these boundary integrals.   

Special care has to be taken at the contact point. In other studies \citep{vandre2013mechanism,vandre2012delay,liu2019predictions,liu2016assist,liu2016surfactant,liu2017mechanism} the contact angle is imposed as an essential boundary condition at the expense of solving a component of the momentum equations at the contact point. We adopt the approach of \cite{sprittles2013finite} and impose the contact angle as a natural boundary condition on both the intersection of the free-surface with the left plate ($\Gamma_1$) and the symmetry plate ($\Gamma_2$). We therefore introduce a field of Lagrange multiplier unknowns on $\Gamma_1$ and $\Gamma_2$ which are determined from the weak form of the no-penetration condition. We refer the reader to \cite{sprittles2013finite} for a detailed description of this implementation (we adopt approach (B) in their nomenclature).

As is standard, the fluid velocities are interpolated using bi-quadratic shape functions and the pressure using linear continuous shape functions with Taylor-Hood triangular elements. We choose to mesh the liquid domain using an unstructured triangular grid; see figures~\ref{fig:domain}(b) and (c). The mesh is considered to be a fictitious pseudo-solid with the position of the nodes coming as part of the solution. The weak form of the kinematic condition, \eqref{kinematic_condition}, is imposed as an essential condition and determines a field of Lagrange multipliers (not to be confused with the Lagrange multipliers in the previous paragraph) that act on the solid deformation equations which in turn determines the shape of the unknown interface, $\textbf{r}$, see \cite{sackinger1996newton} for more details. We note that this approach results in a large system of equations which is disadvantageous, but it also allows for the interface to become highly deformed and even multi-valued, as well as naturally handling unsteady flow (where the domain could significantly change shape c.f. Figure~\ref{fig:system_sketch}), which is difficult to achieve if the mesh is structured.

To solve the hybrid equation, \eqref{hybrid_equation}, it is convenient to introduce two fields of unknowns on the fluid interface, the pressure $p_1$ and flux $Q_1$, interpolated using quadratic shape functions. We solve two equations in their weak form:
\bea
\int_{S}(Q_1 - Q_{C})\psi\,\mbox{d}S = 0,\qquad Q_{C} = \frac{1}{\chi}\left(Ah + \frac{1}{2}Bh^2 + \frac{1}{6}h^3\pdiff{p}{s}\right)
\label{hybridflux}
\eea
and
\bea
\int_{S} \left(\pdiff{h}{t} + \pdiff{Q_1}{s}\right)\psi\,\mbox{d}s = 0
\label{hybridmass}
\eea
Equation \eqref{hybridflux} projects the flux from the lubrication equation onto the finite element space and then \eqref{hybridmass} ensures mass is conserved in the fluid phase.

The resulting discretised equations are solved using Newton's method using the SuperLu numerical algebra package \citep{li2005overview}. For time-dependent calculations the solution is updated in time using a backwards-difference second-order Euler method (BDF2). 

Around the contact line the interface becomes highly deformed due to viscous bending and the pressure and velocity gradients are large, see figures~\ref{fig:domain}(d) and (e). In steady calculations, as $Ca\to Ca_{\mathrm{crit}}$, we expect the number of elements required in the vicinity of the contact line to increase to ensure a smooth converged solution. We re-mesh the domain according to a ZZ error estimator \citep{Zienkiewicz1992}, which measures the discontinuity of strain rate gradients between adjacent elements and interprets this as a measure for the local error. Typically we set a minimum error as $10^{-6}$ and a maximum error $10^{-3}$, so that elements with error above this range get refined and those with error below this range get unrefined. We allow element sizes from $10^{-12}$ to $10^{-2}$ to accommodate these error estimates. We do not adapt the mesh at each calculation; rather we adapt the mesh based on the condition that
\bea
|\theta_1 - \theta_c| < 1.0^{\circ},\qquad \theta_c = \mbox{atan} ((y_2 - y_1)/(x_2 - x_1))
\eea
where $\theta_c$ is the computed angle based on $(x_1,y_1)$ and $(x_2,y_2)$, the coordinates of the nodes on $\Gamma_4$ directly at the contact line and immediately adjacent, respectively. The number of elements and their sizes are highly dependent on $\lambda$ and $Ca$. As an illustrative example, for steady solutions at $\lambda = 0.1$, $\chi=0.1$ and $V=5$, the resulting mesh has $\sim 10^3$ triangular elements and $\sim 10^5$ discretised unknowns at $Ca = Ca_{\mathrm{crit}}$.

\section{Linear Stability Analysis} \label{sec:stability}

We now present the stability algorithm and results. Rather than perform a standard normal modes reduction to the Orr-Somerfeld equations - \citep[see, for example][]{severtson1996stability} we take a more general approach that determines the modes as part of the solution. The analysis below is independent of the model, and although the results we present are from the hybrid model, these results also follow from the full model.

In both cases the PDE system can be written as
\bea
\mathcal{R}(\dot{\textbf{w}},\textbf{w}) = 0,
\label{nonlinear_system}
\eea
where $\mathcal{R}$ is a nonlinear operator and $\textbf{w}(t)$ represents a state of the system at time $t$, given as a vector of all the unknowns (either \eqref{full_unknowns} or \eqref{hybrid_unknowns}). The time derivatives, $\dot{\textbf{w}}$, appear in linear combinations in our system so we can decompose $\mathcal{R}$ into a linear mass operator, $\mathcal{M}$, that operates on the time-derivatives in the problem and a nonlinear operator, $\mathcal{F}$, that operates on the spatial derivatives in the problem so that \eqref{nonlinear_system} becomes
\bea
\mathcal{R}(\dot{\textbf{w}},\textbf{w}) \equiv \mathcal{M}(\dot{\textbf{w}}) + \mathcal{F}(\textbf{w}) = 0.
\label{master_eqn}
\eea

To proceed we write the state of the system, $\textbf{w}$, as a Taylor expansion, i.e.
\bea
\textbf{w} = \textbf{w}_{\star} + \varepsilon \mbox{e}^{\sigma t}\textbf{g} + O(\varepsilon^2),
\label{perturbation}
\eea
where $\textbf{w}_{\star}$ is a base state only dependent on spatial variables, $\varepsilon\ll 1$ is a small parameter, $\textbf{g}$ is an eigenmode that is dependent on spatial variables only and $\sigma$ is the growth rate of the perturbation. The expansion in \eqref{perturbation} represents a general class of perturbations that satisfy the boundary conditions of the problem and are in-plane perturbations; we are not extending to the third dimension, a problem we will discuss later.

Substituting \eqref{perturbation} into \eqref{master_eqn} gives a series of problems that have to be solved at each order of $\varepsilon$. At leading order we have
\bea
\mathcal{F}(\textbf{w}_{\star}) = 0.
\label{steady_eqn}
\eea
The solution, $\textbf{w}_{\star}$, is the steady state of the system. At first order we solve
\bea
\left[\sigma\mathcal{M}(\textbf{w}_{\star}) + \mathcal{J}(\textbf{w}_{\star})\right]\textbf{g} = 0,
\label{eigen_eqn}
\eea
where $\mathcal{J}(\textbf{w}_{\star})$ is the functional derivative of the nonlinear operator $\mathcal{F}$ applied at the steady state $\textbf{w}=\textbf{w}_{\star}$. Equation \eqref{eigen_eqn} is a generalised eigenvalue problem that can be solved to find $\textbf{g}$ and $\sigma$. The eigenspectrum of $\sigma$ determines the stability of the steady solutions. If at least one of the spectrum of $\sigma$ has a positive real part then the steady state is linearly unstable. Conversely if the entire spectrum lies in the left-half of the complex plane then the solution is stable. In general there will be an infinite number of these eigenmodes and thus we can write the linearised solution as
\bea
\textbf{w}(t) = \textbf{w}_{\star} + \sum_{n=1}^{\infty}a_n\textbf{g}_n\mbox{e}^{\sigma_nt} + \mbox{c.c},
\label{eigenmode_sum}
\eea
where c.c denotes the complex conjugate and $a_n$ are arbitrary constants set by the initial conditions. When the system becomes discretised the operators $\mathcal{M}$ and $\mathcal{J}$ are represented by the mass-matrix and Jacobian matrix, respectively. The mass matrix representation of $\mathcal{M}$ is highly rank deficient as the only time-derivatives occur at the fluid-liquid interface and special care has to be taken to ensure that the solution to \eqref{eigen_eqn} has converged. We use the Anazasi linear algebra library which is an iterative eigensolver that can solve highly rank-deficient eigenproblems \citep{herouxtrilnos}. As the spectrum has an infinite number eigenvalues, the discretised spectrum will have a finite number of eigenvalues, proportional to the number of unknowns in the problem. We find a small subset of eigenvalues which have the largest real part as these will be the modes visible in the transient dynamics; large negative eigenvalues correspond to eigenmodes that decay very rapidly. We validate the calculations using a simplified lubrication model and present this in Appendix~\ref{sec:eigenmode_validation}.

\subsection{Stability of the Solution Branches}
\begin{figure}
\centering
  \includegraphics[scale=0.26,trim=0 0 0 0,clip]{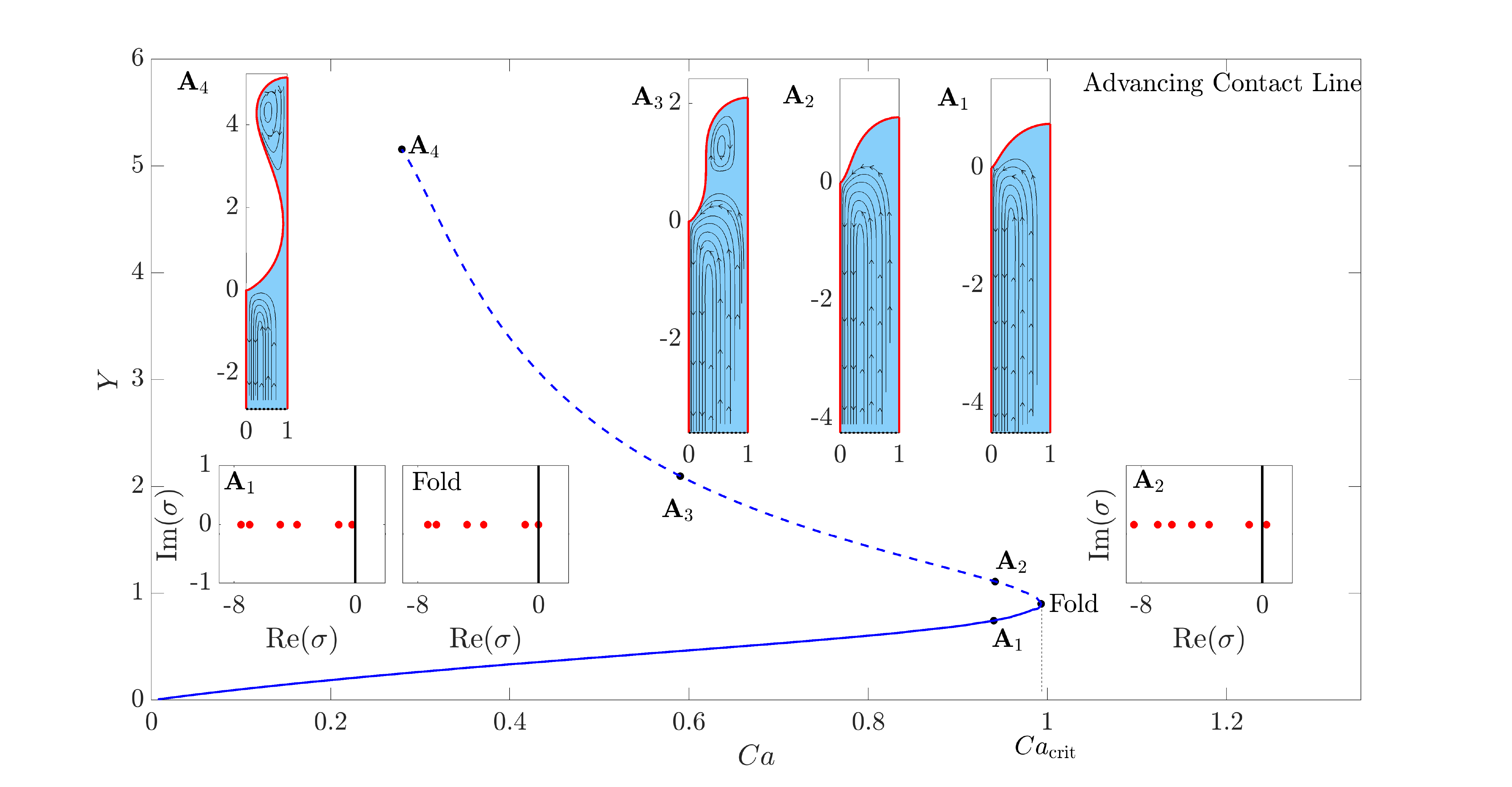}
  \caption{Solution curve for the advancing contact line. The values of the control parameters are $\lambda = 0.1,\chi = 0.1, V = 5.0$. Individual solutions are labelled on the curve and correspond to the inset panels with the same label. The curve indicated by solid/dashed lines are stable/unstable, respectively. The eigenspectra for the solutions either side of the fold, for solutions \textbf{A}\textsubscript{1} and \textbf{A}\textsubscript{2}. \textbf{A}\textsubscript{3} is the solution where the inflection point on the interface first becomes parallel to the wall, i.e. when $\mbox{d}x/\mbox{d}y = 0$. \textbf{A}\textsubscript{4} is the solution just before the numerical calculations cease to converge which occurs when the interface becomes sufficiently deformed so that it touches the left plate.}
  \label{fig:advancing_bif_curve}
\end{figure}
\begin{figure}
\centering
  \includegraphics[scale=0.30,trim = 0 0 0 0,clip]{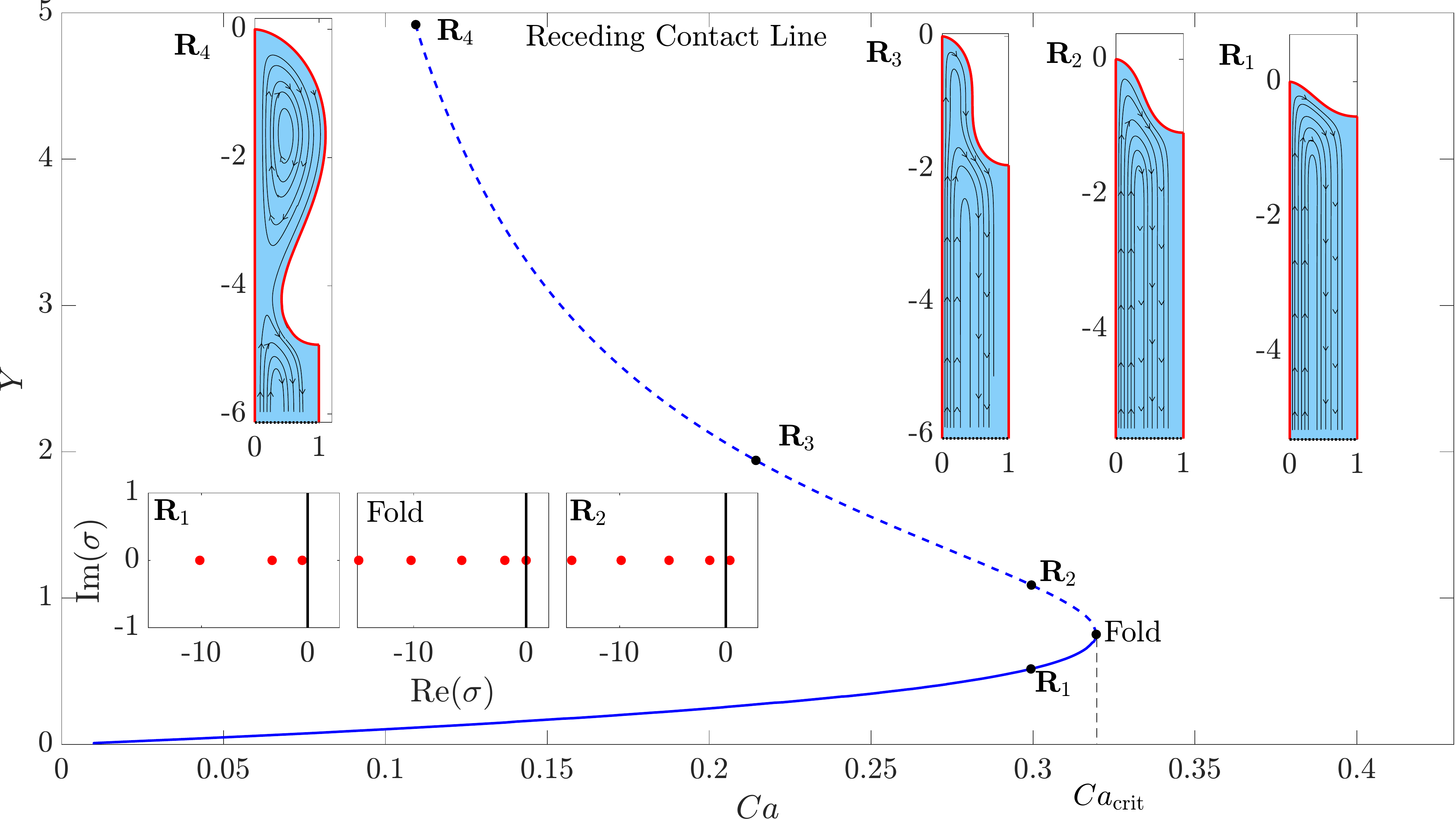}
  \caption{Solution curve for the receding contact line. The values of the control parameters are $\lambda = 0.1,\chi = 0.0, V = 5.0$. Individual solutions are labelled on the curve and correspond to the inset panels with the same label. The curve indicated by solid/dashed lines are stable/unstable, respectively. The eigenspectra for the solutions either side of the fold, for solutions \textbf{R}\textsubscript{1} and \textbf{R}\textsubscript{2}. \textbf{R}\textsubscript{3} is the solution where the inflection point on the interface first becomes stationary, i.e. when $\mbox{d}x/\mbox{d}y = 0$. \textbf{R}\textsubscript{4} is the solution just before numerical calculations cease to converge which occurs when the interface becomes sufficiently deformed so that it touches the stationary right plate}
  \label{fig:receding_bif_curve}
\end{figure}
\begin{figure}
\centering
	\includegraphics[scale=0.25]{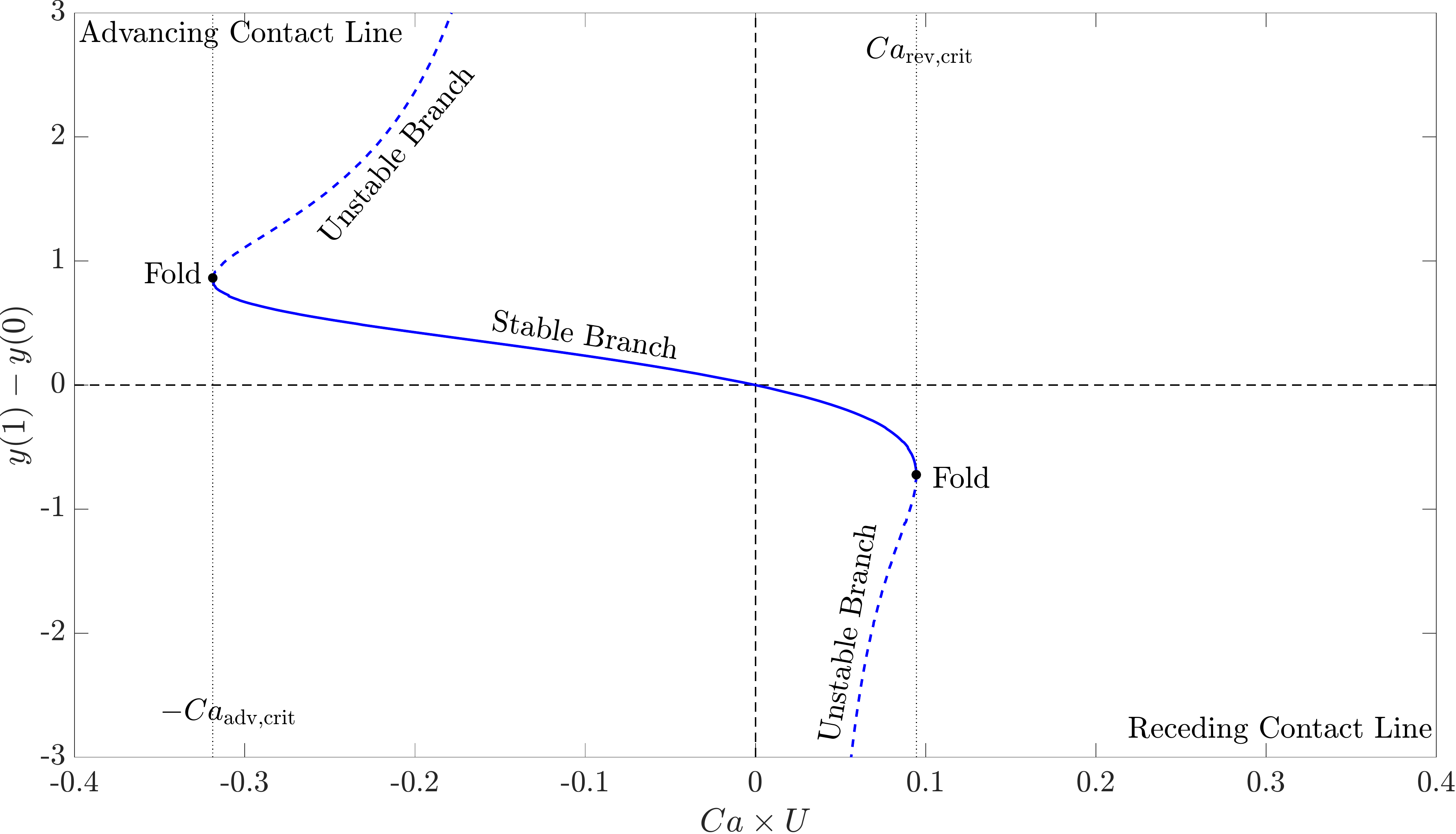}
	\caption{The advancing and receding steady solution curves for $\lambda = 0.01,\chi = 0.1$. The horizontal axis is $Ca$ scaled by $U=\pm 1$ and the vertical axis is $y(1) - y(0)$, i.e. the signed version of $Y$. The critical $Ca$ for each problem is denoted by a dotted line. The two problems are connected through the origin.}
	\label{fig:advancing_receding_comparison}
\end{figure}

We now discuss the bifurcation structure and the corresponding stability results of the advancing and receding dynamic contact line problems. Figures~\ref{fig:advancing_bif_curve} and \ref{fig:receding_bif_curve} show the bifurcation structures in a typical advancing case ($\chi = 0.1, \lambda = 0.1$) and receding case ($\chi = 0.0, \lambda = 0.1$) respectively. Notably, our focus here is on providing insight into the stability structure, rather than necessarily probing the precise values from experimental analyses, where the slip length could be far smaller and therefore typically require more computational resources. Previous works \citep{vandre2012delay} have shown that whilst changes in slip length can have a weak effect on $Ca_{\mathrm{crit}}$, they do not qualitatively alter the physical mechanisms at play (similarly for smaller viscosity ratios, e.g. with a glycerol-air system).

The solution curves are shown in the $(Ca,Y)$ projection of the solution space (see \eqref{delta} for a definition of $Y$). The markers on the curve indicate specific solutions which are shown in the inset panels labelled \textbf{A}\textsubscript{1}-\textbf{A}\textsubscript{4} for the advancing contact line, and \textbf{R}\textsubscript{1}-\textbf{R}\textsubscript{4} for the receding contact line. The eigenspectra of \textbf{A}\textsubscript{1},\textbf{A}\textsubscript{2} and \textbf{R}\textsubscript{1},\textbf{R}\textsubscript{2}, and at the fold are shown in inset panels for the advancing and receding contact lines respectively. In both the advancing and receding cases, as $Y$ increases, the solution curve experiences a fold which separates the lower branch and upper branch. The eigenspectra is real and at the fold a single eigenvalue crosses the imaginary axis, as expected. The eigenspectra also indicates that the \textbf{A}\textsubscript{1}/\textbf{R}\textsubscript{1} states are `attractors' of the system and \textbf{A}\textsubscript{2}/\textbf{R}\textsubscript{2} states are weakly unstable `saddle-nodes' (figure~\ref{fig:bifurcation_sketch}(b)), thus numerically confirming the lower branch is stable (solid curve) and the upper branch is unstable (dashed curve).

The interface has an inflection point near the contact point. We measure the angle at the interface inflection point (to the downwards vertical) and define this as $\theta_{\mathrm{app}}$; the apparent contact angle \citep{vandre2012delay,liu2016surfactant}. Notably, as can be seen from solutions \textbf{A}\textsubscript{1},\textbf{A}\textsubscript{2} and \textbf{R}\textsubscript{1},\textbf{R}\textsubscript{2} in figures~\ref{fig:advancing_bif_curve} and \ref{fig:receding_bif_curve}, $\theta_{\mathrm{app}}<180^\circ$ not only on the stable branch, but also immediately after the fold on the unstable branch. Further up the unstable branch, as can be seen from solutions \textbf{A}\textsubscript{3},\textbf{A}\textsubscript{4} and \textbf{R}\textsubscript{3},\textbf{R}\textsubscript{4} in figures~\ref{fig:advancing_bif_curve} and \ref{fig:receding_bif_curve}, $\theta_{\mathrm{app}} \geq 180^{\circ}$ and the interface becomes multi-valued (as a function of $h$). In the advancing/receding cases the solution curve terminates when the interface is sufficiently deformed so that the interface touches the right/left plate, respectively, so that the interface effectively `pinches' off the fluid domain, as seen in solutions \textbf{A}\textsubscript{4} and \textbf{R}\textsubscript{4} in figures~\ref{fig:advancing_bif_curve} and ~\ref{fig:receding_bif_curve}. 

The steady-solution curves of the advancing and receding contact line problems, although treated separately in figures~\ref{fig:advancing_bif_curve} and \ref{fig:receding_bif_curve}, are actually two halves of the same solution space. Figure~\ref{fig:advancing_receding_comparison} shows the connection for $\lambda = 0.01,\chi = 0.1,\theta_m = \pi/2$ where the signed meniscus rise, $y(1) - y(0)$, is plotted against $Ca\times U$, where $U=\pm 1$ with $+/-$ corresponding to the receding/advancing problem respectively. The advancing and receding curves occupy the second and fourth quadrants in this projection and the location of the respective folds in each quadrant highlights that the receding contact line becomes unstable before the advancing contact line \citep{chan2013hydrodynamics}. 

Finally, we note that in both cases the solution curve does not experience additional bifurcations as $Y$ increases along the unstable branch. For a system where gravity is included it is known that within the lubrication approximation, the solution curve (for the receding contact line, at least) oscillates around a fixed value of $Ca = Ca^*$, \citep[see][]{snoeijer2012theory}, experiencing multiple saddle-node bifurcations as $Y\to \infty$. Preliminary calculations show that, if gravity is included, the oscillations are also present in the advancing/receding hybrid system, although, for brevity, we do not show the results here. 

\subsection{Physical Interpretation of the Bifurcation}  
\begin{figure}
    \centering
    \includegraphics[scale=0.25]{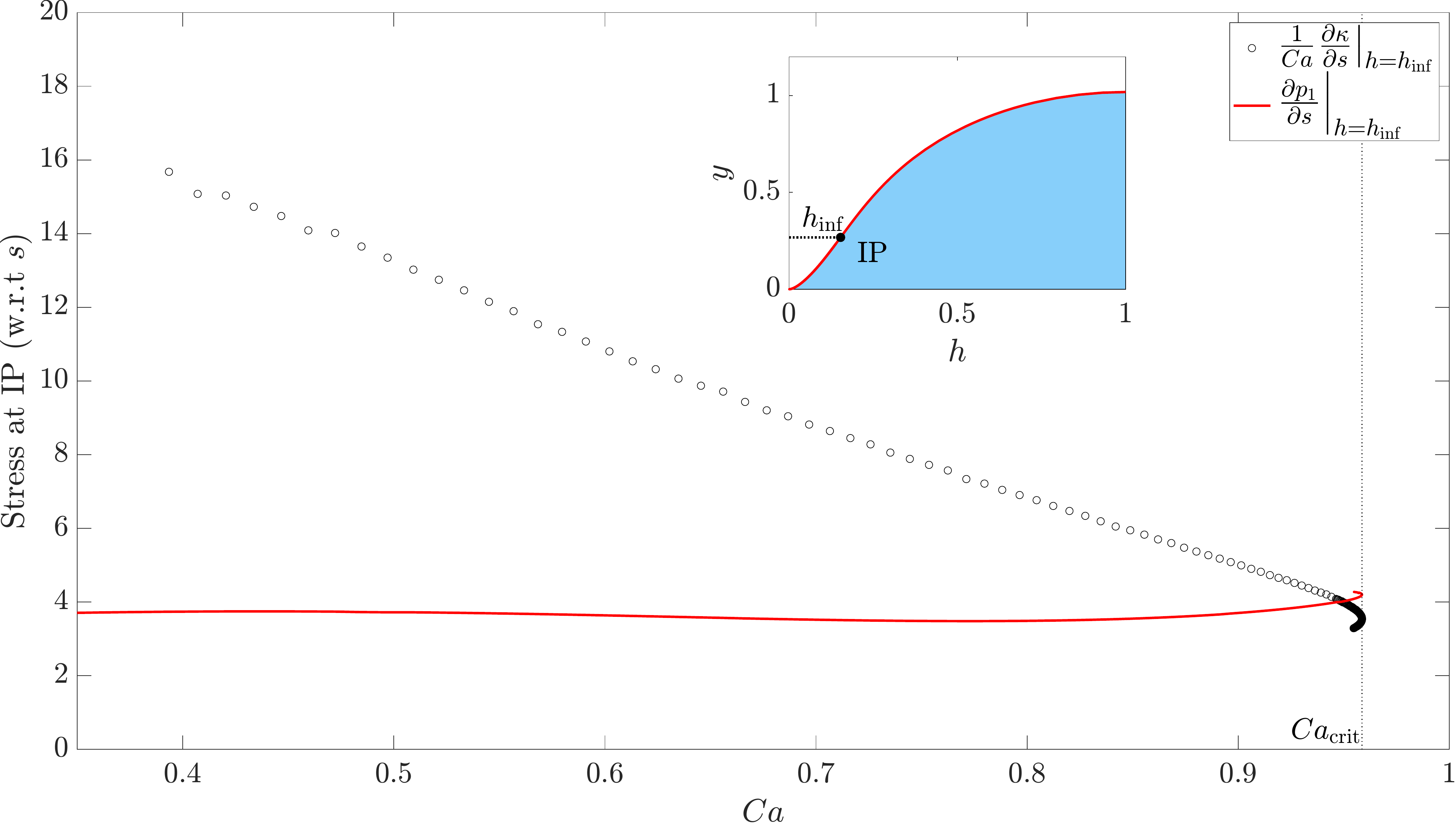}
	  \caption{Comparison of the air pressure gradient and the capillary stress gradient for the steady solutions at the inflection point on the interface for the advancing contact line problem. Parameter values are $\chi=0.1,\lambda = 0.1$.}
    \label{fig:stress_gradients}
  \end{figure}
As discussed in \cite{vandre2013mechanism} for the advancing contact line, the fold occurs when the fluid pressure gradients (fluid 1) are comparable to the capillary stress gradients \citep[see][]{vandrethesis} near the contact line, i.e. when
  \bea
  \diff{p_1}{s} \sim \frac{1}{Ca}\diff{\kappa}{s}.
\label{stress}
\eea
This is because as $Ca\to Ca_{\mathrm{crit}}$ the air pressure gradients near the contact-line will increase as the system seeks to `pump' air out of the region near the contact line to maintain system state. Eventually these air pressure gradients will exceed the capillary stress gradient and the system will be unable to maintain a stable steady equilibrium. Figure~\ref{fig:stress_gradients} shows the evolution of the quantities on either side of \eqref{stress} calculated at the inflection point for the advancing case. The pressure and capillary stress gradients balance close to $Ca_{\mathrm{crit}}$, as seen by the intersection of the curves, which confirms the ideas of \cite{vandre2013mechanism}.

\subsection{Fold-Tracking}
\begin{figure}
\centering
  \includegraphics[scale=0.3]{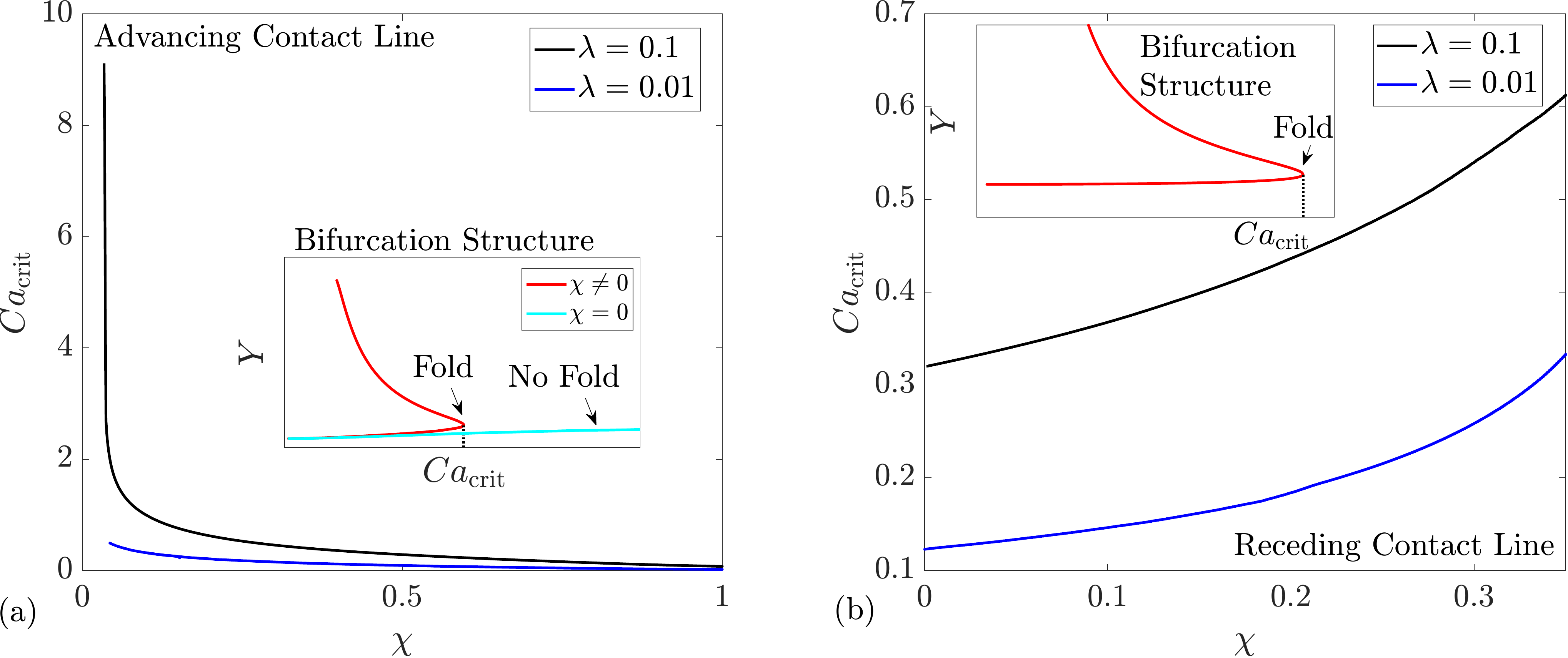}
  \caption{The evolution of the fold, and hence $Ca_{\mathrm{crit}}$, as the relative viscosity, $\chi$, of the fluid to liquid phase is varied for $\lambda = 0.01,0.1$.  (a) The advancing contact line. As $\chi\to 0$, $Ca_{\mathrm{crit}}\to\infty$ so that the fold in the bifurcation structure ceases to exist. The inset shows the generic bifurcation structure in the $\chi=0$ and $\chi\neq 0$ cases. In the former case there is always a stable solution for the system to be attracted to. (b) The receding contact line. The fold exists for all viscosity ratios. For a given $\chi$ the bifurcation structure is shown in the inset. We note the qualitative behaviour shown in the inset is independent of the value of $\lambda$.}
  \label{fig:ca_crit_evolution}
\end{figure}
We can take advantage of the fact that at the fold bifurcation the leading eigenvalue crosses the imaginary axis to develop an algorithm for finding $Ca_{\mathrm{crit}}$. We augment the system with the additional constraint
\bea
\Real(\sigma_1) = 0,
\label{fold_eqn}
\eea
and let another control parameter come as part of the solution. It is convenient to let the interface length, $L$, be determined by \eqref{fold_eqn} so we are able to track the evolution of the $Ca_{\mathrm{crit}}$ as another parameter, the viscosity ratio $\chi$, for example, is varied. This is a robust way of tracking the fold without having to recalculate the solution curve for every set of parameters, as previously considered in \cite{kamal2019dynamic} and \cite{vandre2012delay}.

If we vary $\chi$ and calculate $Ca_{\mathrm{crit}}$ we observe that the curve of the loci of $Ca_{\mathrm{crit}}$ does not itself experience any bifurcation, (co-dimension 2 bifurcations), as seen in figures~\ref{fig:ca_crit_evolution}(a) and (b). In addition we observe that that the bifurcation structure also remains intact when the slip-length, $\lambda$, is varied as the different coloured curves in figures~\ref{fig:ca_crit_evolution}(a) and (b) indicate. Therefore we expect the dynamics to be qualitatively similar (from a dynamical systems perspective) regardless of the viscosity ratio or slip-length.

An important observation is that the advancing and receding cases differ significantly as $\chi\to 0$. For the advancing case, $Ca_{\mathrm{crit}}\to\infty$ in this limit, whilst for the receding case it tends to a finite value. This indicates that the viscosity of the fluid phase has to be taken into account for the advancing contact line in order to describe the bifurcation structure. In contrast the receding contact line is essentially a one-phase problem, particularly if the fluid is a gas and qualitative features of the bifurcation structure are the same regardless of the viscosity of the fluid.

\subsection{Eigenmode Perturbations}

\begin{figure}
  \centering
\includegraphics[scale=0.3]{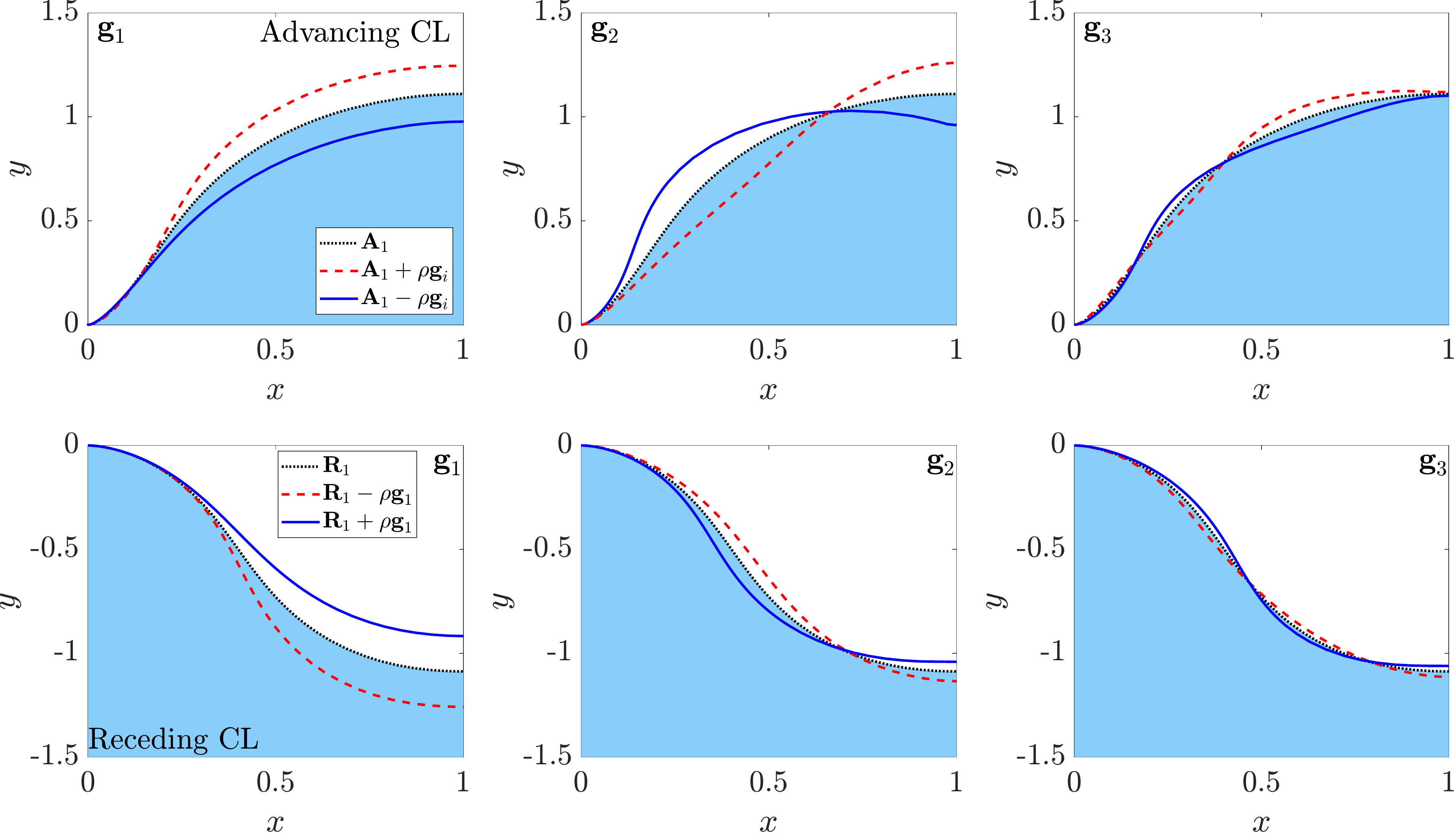}
\caption{The eigenmodes of the unstable branch. The steady profile is shown as the dotted line. Top row: The leading eigenmodes, $\textbf{g}_1,\textbf{g}_2,\textbf{g}_3$, corresponding to the advancing $\textbf{w}_{\star} = \textbf{A}_2$ ($Ca=0.934,\chi=0.1$) solution infigure~\ref{fig:advancing_bif_curve}. Bottom row: The leading eigenmodes corresponding to the receding $\textbf{w}_{\star} = \textbf{R}_2$ ($Ca=0.3,\chi=0$) solution in figure~\ref{fig:advancing_bif_curve}. In each case the eigenmodes cross the steady interface in successively more locations. The slip length is $\lambda = 0.1$.}
\label{fig:eigenmodes}
\end{figure}

We now discuss the nature of the eigenmodes resulting from the stability analysis. The modes corresponding to the three leading eigenvalues of the unstable branch are shown in figure~\ref{fig:eigenmodes}. These eigenmodes correspond to the base state $\textbf{w}_{\star} = \textbf{A}_2,\textbf{R}_2$ in figures~\ref{fig:advancing_bif_curve} and \ref{fig:receding_bif_curve}, respectively. In this figure the dotted profile indicates the steady interface shape and the coloured lines indicate the shape of the interface when it is perturbed by a single eigenmode, i.e.
\bea
\textbf{w} = \textbf{w}_{\star} \pm \rho \textbf{g}_i,\qquad i = 1,2,3.
\label{stable_pert}
\eea
The dashed/solid curves correspond to the $+/-$ sign, respectively. The {amplitude} of the perturbation, $\rho$, is constrained so that the meniscus rise of the perturbation is no more than 10\% of the rise of the steady solution. Each successive mode intersects the steady interface at precisely one more location in similarity to the form of the eigenmodes in a related lubrication model (see figure~\ref{fig:eigenmode_validation}). Thus, the effect of adding higher-order eigenmodes to the steady state is add extra corrugations to the interface. 
  \begin{figure}
    \centering
    \includegraphics[scale = 0.3]{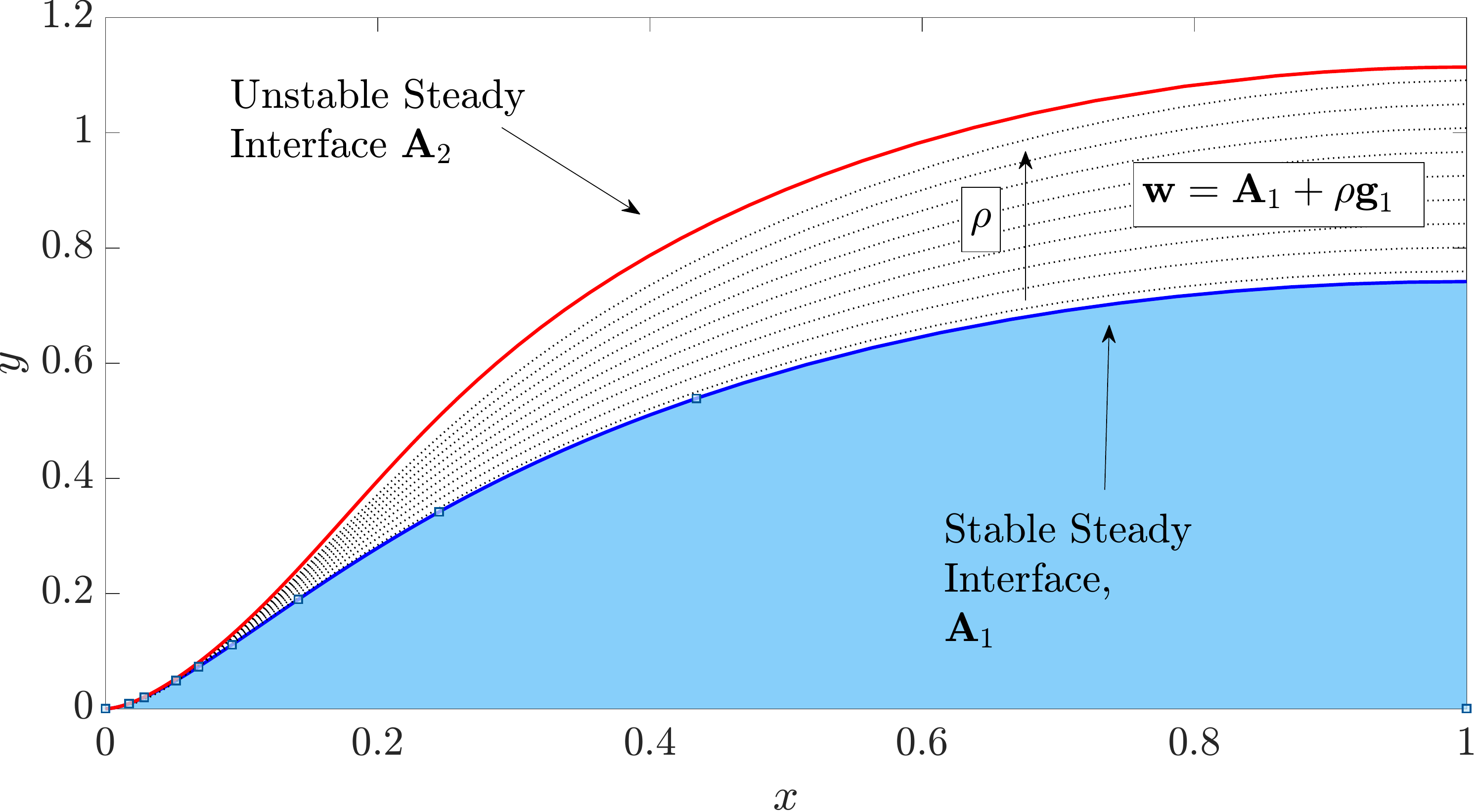}
    \caption{Perturbations of the advancing $\textbf{w}_{\star}=\textbf{A}_1$ state at $Ca=0.934$. The lower solid curve is the stable interface and the upper solid curve is the unstable interface. The dotted curves are different strength perturbations of the stable state according to \eqref{stable_pert} with $i=1$. Parameter values are $Ca = 0.934,\chi=0.1,\lambda = 0.1$.}
    \label{fig:stable_pert_interface}
  \end{figure}
\begin{figure}
  \centering
\includegraphics[scale=0.3,trim= 0 0 0 0,clip]{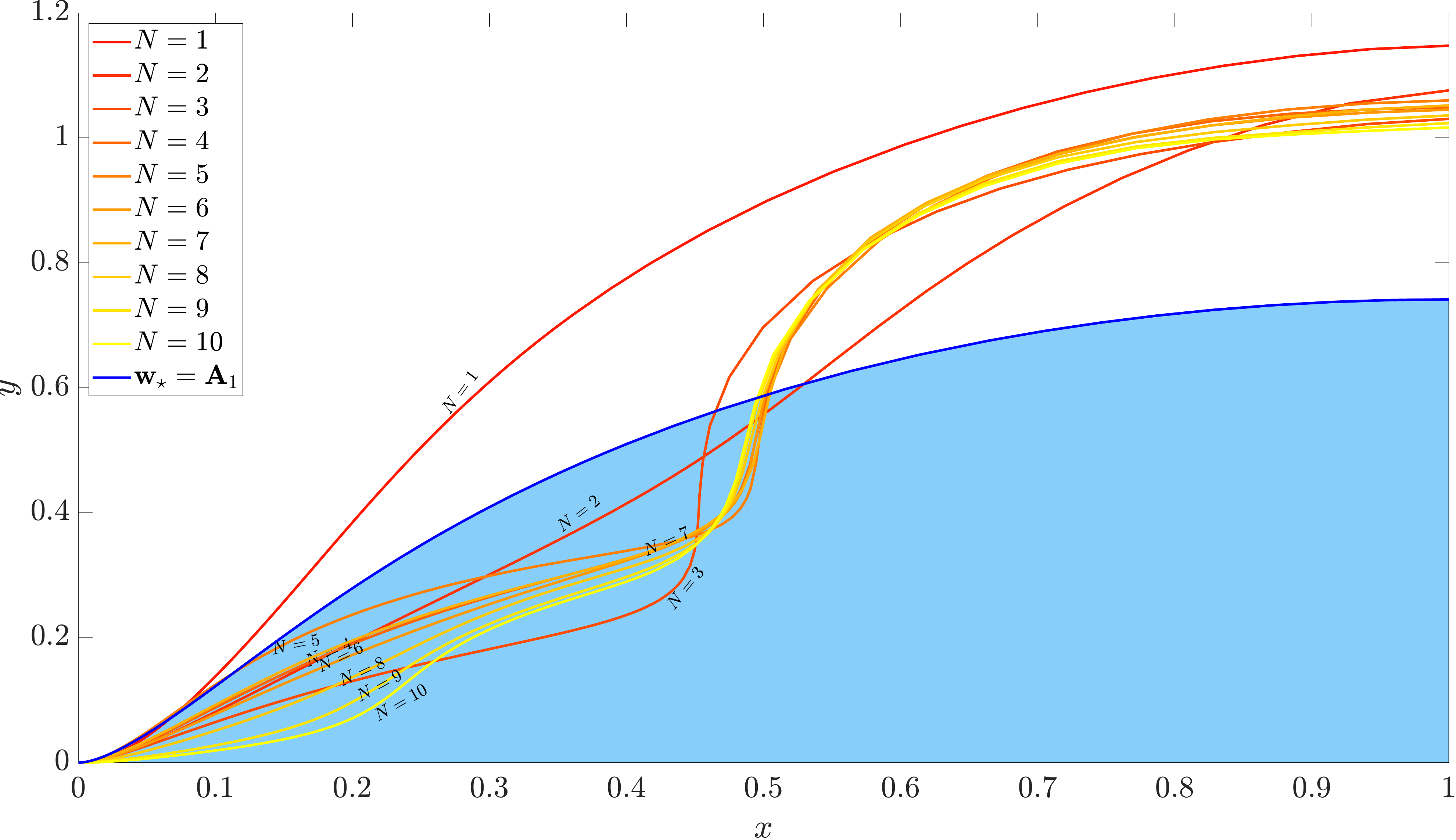}
	\caption{Perturbations of the advancing $\textbf{w}_{\star} = \textbf{A}_1$ state using \eqref{mixed_stable_pert} for a fixed value of $\rho$ and increasing $N$. The shaded region represents the nonlinear \textbf{A}\textsubscript{1} steady solution and the coloured curves are the perturbed states using \eqref{mixed_stable_pert}. Parameter values are $Ca = 0.934,\chi=0.1,\lambda = 0.1$.}
	\label{fig:increasing_N}
\end{figure}

Concentrating on the leading eigenmode alone, the action of adding a multiple of $\textbf{g}_1$ to a steady solution, i.e. taking $i=1$ in \eqref{stable_pert}, stretches/shrinks the interface according to the $\pm$ sign with no additional corrugations. Figure~\ref{fig:stable_pert_interface} shows the stable \textbf{A}\textsubscript{1}, and unstable \textbf{A}\textsubscript{2} states with solid lines and the dotted curves indicate the perturbed interface profiles from the \textbf{A}\textsubscript{1} state using the leading $\textbf{g}_1$ eigenmode. This figure demonstrates that if we can continuously `stretch' the nonlinear stable state by increasing the strength of the perturbation, $\rho$, and can eventually achieve an interface profile similar to the unstable steady state, \textbf{A}\textsubscript{2}, which will have consequences, as discussed in the next section. We can also continuously deform the unstable branch in the same manner to match the interface of the stable branch. We shall denote perturbations using the leading eigenmode only in \eqref{stable_pert} as stretch perturbations.

In a physical experiment, perturbations will naturally emerge from the presence of `noise' in the system. This `noise' will occur from random fluctuations of the interface and contact-line, but rather than apply a stochastic perturbation to the system we will mimic this `noise' by using more eigenmodes in the perturbation, i.e. 
 \bea
\textbf{w} = \textbf{w}_{\star} + \sum_{i=1}^{N}\rho_i\textbf{g}_i,
\label{mixed_stable_pert}
\eea
where for the purpose of simple illustration we choose the amplitude coefficients, $\rho_i$, to be equal. Figure~\ref{fig:increasing_N} shows the perturbed interface as the value of $N$ increases from 1 to 10, which demonstrates that by increasing the value of $N$ we are able to perturb the nonlinear steady interface with increasingly more corrugations or `noise'. Henceforth we shall call perturbations of this form as corrugation perturbations.

The two forms of perturbation discussed here, namely the `stretch' and `corrugation' perturbations, will be used in the next section to understand the subsequent time-dependent behaviour of the system after systematically applying a perturbation to a steady state.
  
\section{Transient Dynamics}\label{sec:transient_dynamics}
Now that we have calculated the steady solution branches and quantified their stability, we attempt to answer the two questions of fundamental importance:

\begin{enumerate}
  \item How do the steady-states, stable and unstable, and the resulting bifurcation structure help us understand the time-dependent behaviour of the system when $Ca<Ca_{\mathrm{crit}}$? 
\item What is the time-dependent behaviour of the system when we choose initial conditions (IC) beyond the fold, i.e. when $Ca>Ca_{\mathrm{crit}}$?
\end{enumerate}

To address these questions we solve the time-dependent hybrid PDE as an IVP. It is useful to define solution measures that will help visualise and aid the discussion. In our formulation $Ca$ corresponds to the speed of the wall and not the speed of the contact point. It is therefore useful to introduce an `effective' $Ca$, based on the contact line speeds relative to the wall (as discussed in \cite{snoeijer2012theory}), which we denote $\overline{Ca}$ and is defined as
\bea
\overline{Ca} = Ca|U - U_{\mathrm{cl}}|,
\eea
where $U=\pm 1$ is the non-dimensional speed of the wall, the $\pm$ sign corresponds to the advancing and receding cases, respectively, and $U_{\mathrm{cl}} = \mbox{d}{y_{\mathrm{cl}}}/\mbox{d}{t}$ is the speed of the contact line. We remark that for steady solutions $U_{\mathrm{cl}} = 0$, and hence $\overline{Ca} = Ca$ and the time-dependent phase-plane trajectories can be directly compared to the bifurcation structure in the $(\overline{Ca},Y)$ plane.

\begin{figure}
    \centering
    \includegraphics[scale = 0.3]{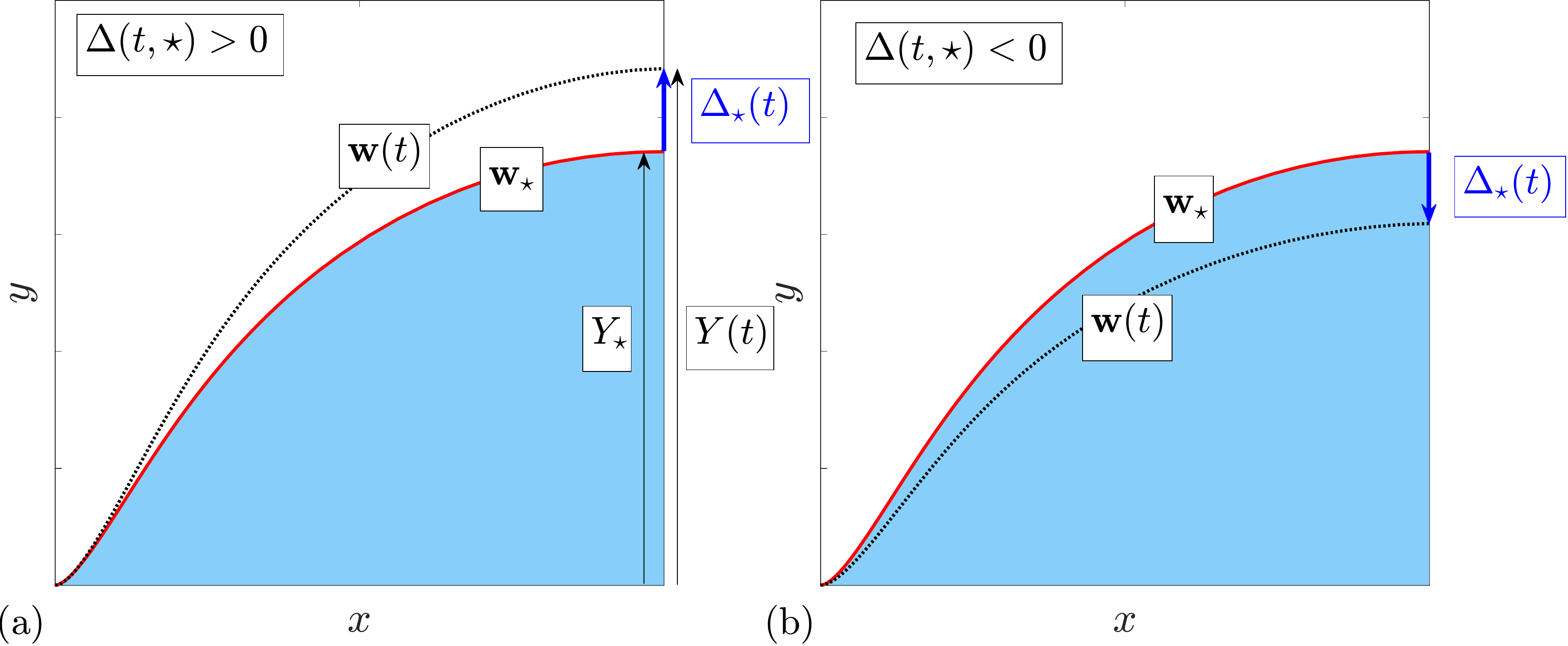}
    \caption{The perturbation measure. The solid curve in each panel is the interface of the base solution, $\textbf{w}_{\star}$, and the dotted curve represents the interface of the system at a given time, $w(t)$. (a) The system is in a state with $\Delta(t,\star) > 0$. (b) The system is in a state with $\Delta(t,\star) < 0$.}
    \label{fig:perturbation_definition}
  \end{figure}

We also introduce a system measure to quantify the size of the perturbation. Let the meniscus rise of a steady solution be $Y_{\star}$, where $\star$ indicates the base solution the perturbation wil be measured against. We can then define a quantity that measures the deviation of the perturbation from the corresponding steady state: 
\bea
\Delta(t,\star) = Y(t) - Y_{\star},
\eea
as demonstrated schematically in figure~\ref{fig:perturbation_definition}. If $\Delta(t,\star)>0$ then the meniscus rise of that current state is larger than that of the steady interface indicated by $\star$ and vice versa if $\Delta(t,\star)< 0$ (see panels (a) and (b) of figure~\ref{fig:perturbation_definition} respectively).
  
\subsection{Perturbations from a steady state IVP: $Ca<Ca_{\mathrm{crit}}$}

We now consider the first question and look at the dynamics of the system when $Ca<Ca_{\mathrm{crit}}$. Our methodology is to start at either the stable or unstable state and perturb it using either a `stretch' or `corrugation' eigenmode expansion, which we will consider separately. We then run a series of IVPs to examine the transient behaviour and eventual dynamical outcome. 

\begin{figure}
  \centering
\includegraphics[scale=0.26,trim=0 0 0 0,clip]{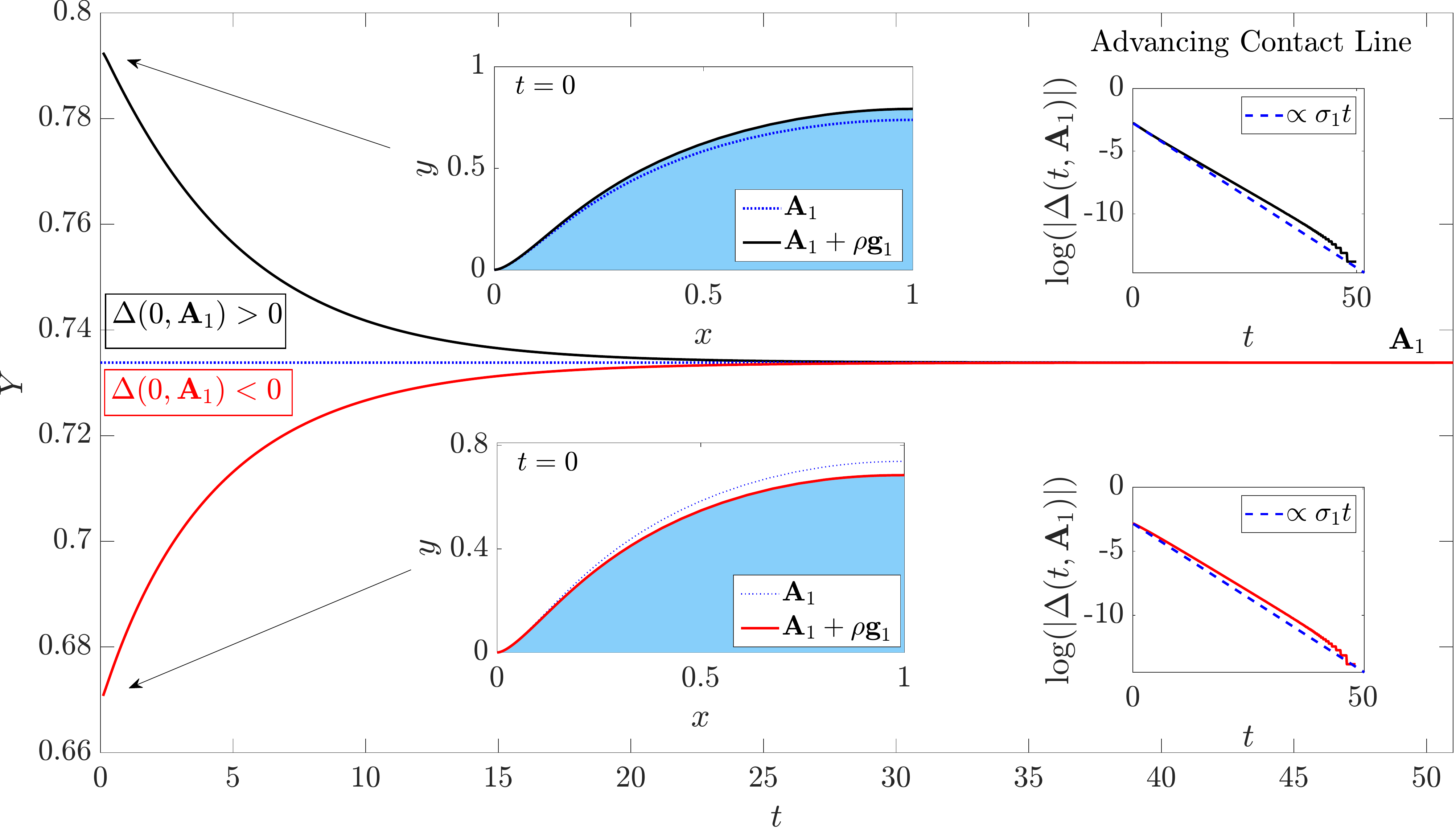}
	\caption{Time-dependent perturbations of the advancing contact line using \eqref{leading_pert}. This figure shows perturbations from the solution labelled \textbf{A}\textsubscript{1} in figure \ref{fig:advancing_bif_curve}. The two different ICs are shown in the inset panels corresponding to the same colour lines in the main panel. Whether $\Delta(0,\textbf{A}_1)>0$ or $\Delta(0,\textbf{A}_1)<0$ the system relaxes back to the stable state. The inset panels on the right show the time signal of $\log(|\Delta(0,\textbf{A}_1)|)$ as compared to the predicted growth rate $\sigma_1$. Parameter values are $Ca = 0.934,\chi=0.1,\lambda = 0.1$.}
\label{fig:advancing_stable_pert}
\end{figure}

\begin{figure}
  \centering
\includegraphics[scale=0.26,trim=0 0 0 0,clip]{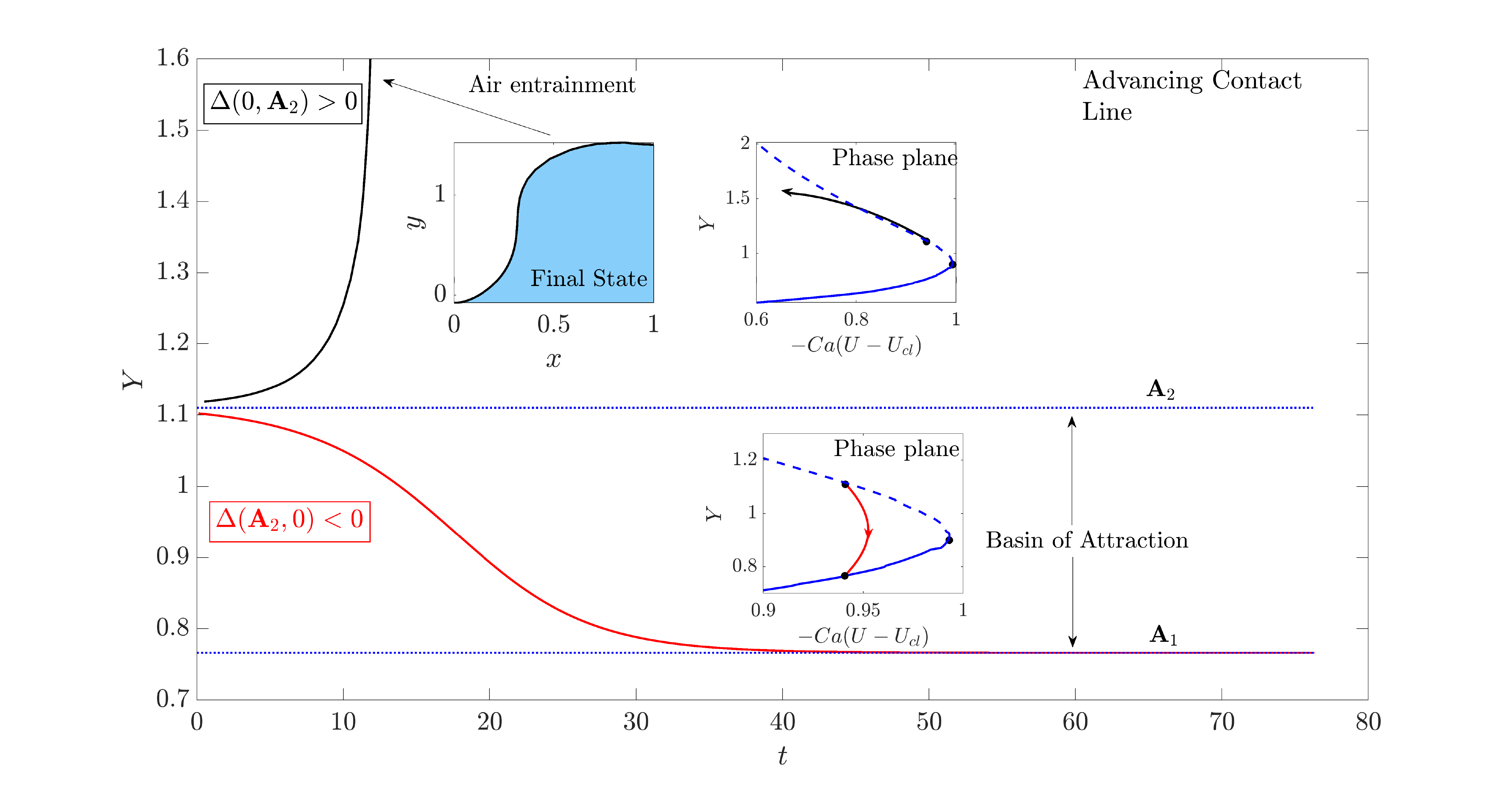}
\caption{Time-dependent perturbations of the advancing contact line using \eqref{leading_pert}. This figure shows perturbations from the solution labelled \textbf{A}\textsubscript{2} in figure \ref{fig:advancing_bif_curve}. When $\Delta(0,\textbf{A}_2)<0$ the perturbation relaxes to the corresponding stable branch. When $\Delta(0,\textbf{A}_2)>0$ the perturbation grows and air entrainment occurs, see inset panel labelled `Final State'. The other inset panels show the phase plane in the $(\overline{Ca},Y)$ projection. The system trajectory is marked with arrows and the steady bifurcation is shown without arrows. The unstable branch can be considered as the basin boundary of attraction of the stable steady state. Parameter values are $Ca = 0.934,\chi=0.1,\lambda = 0.1$.}
\label{fig:advancing_unstable_pert}
\end{figure}

\begin{figure}
  \centering
\includegraphics[scale=0.26]{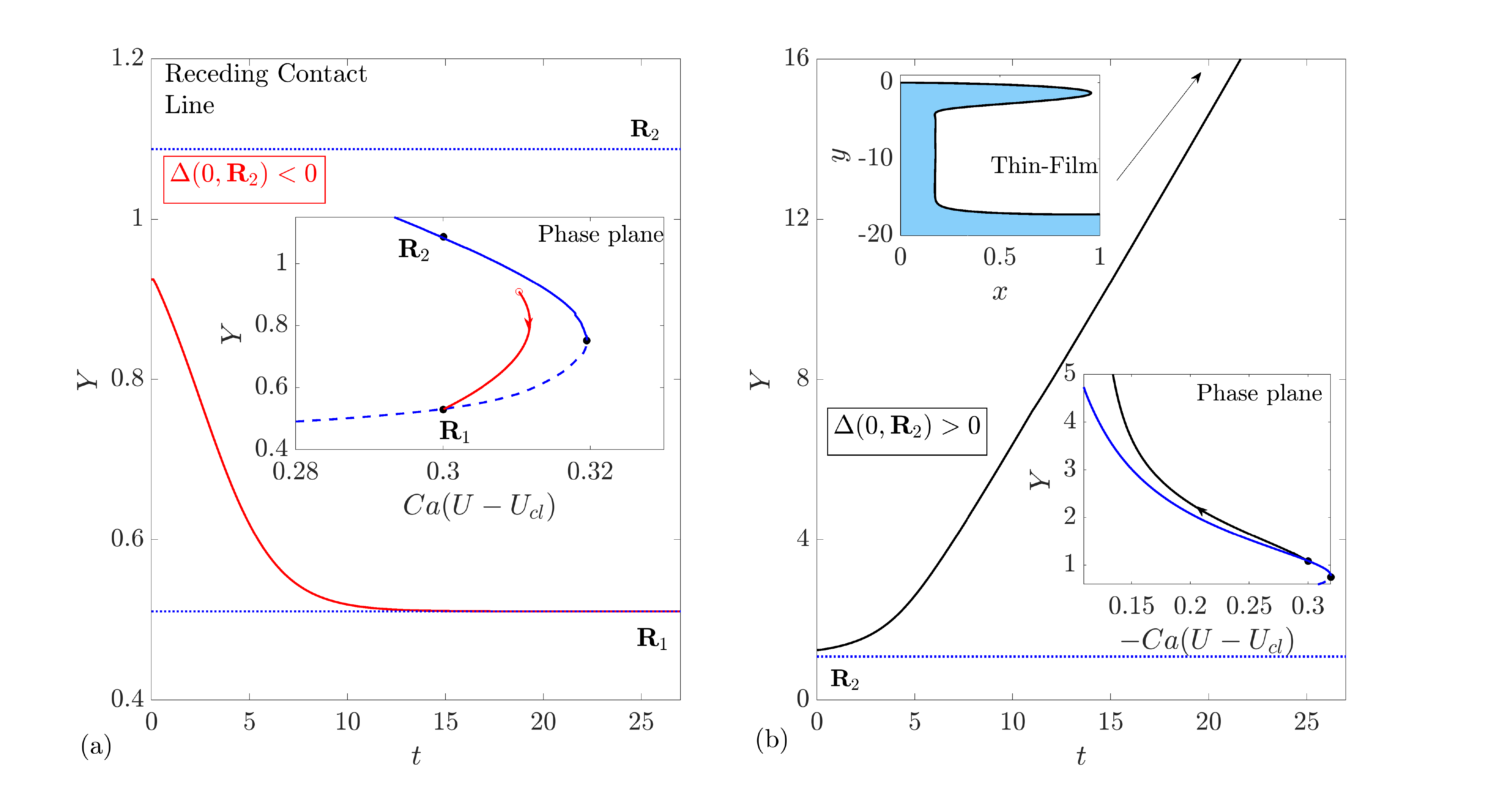}
	\caption{Time-dependent perturbations of the receding contact line unstable branch using \eqref{leading_pert} . This figure shows perturbations from the solution labelled \textbf{R}\textsubscript{2} in \ref{fig:receding_bif_curve}. Panel (a). When $\Delta(0,\textbf{R}_2)<0$ the perturbation relaxes to the corresponding stable branch. Panel (b). When $\Delta(0,\textbf{R}_2)>0$ the perturbation grows and a thin-film develops, see inset panel labelled `Thin-Film'. The other inset panels show the phase plane in the $(\overline{Ca},Y)$ projection. The system trajectory is marked with arrows and the steady bifurcation is shown without arrows. The unstable branch can be considered as the basin boundary of attraction of the stable steady state. Parameter values are $Ca = 0.3,\chi=0,\lambda = 0.1$.}
\label{fig:receding_unstable_pert}
\end{figure}

\subsubsection{`Stretch' perturbations}

For the `stretch' perturbations we use the leading eigenmode only in the perturbation and set the IC to be
\bea
\textbf{w}(t=0) = \textbf{w}_{\star} + \rho \textbf{g}_1.
\label{leading_pert}
\eea
Initially we concentrate on the advancing contact line case. Using the IC stated in \eqref{leading_pert}, small perturbations from the \textbf{A}\textsubscript{1} stable state (figure~\ref{fig:advancing_bif_curve}) decay and the system (unsurprisingly) relaxes back to its stable configuration. Figure~\ref{fig:advancing_stable_pert} demonstrates this by tracking the value of $Y$ in time for two different perturbations, corresponding to $\Delta(0,\textbf{A}_1)<0$ and $\Delta(0,\textbf{A}_1)>0$, of the stable state near the fold (parameter values quoted in the caption). Furthermore, as seen in the insets on the right of figure~\ref{fig:advancing_stable_pert}, the decay rate excellently matches the value of the leading eigenvalue, $\sigma_1$, obtained from the linear stability analysis. For the same parameter values, we can also perturb the \textbf{A}\textsubscript{2} unstable state (figure~\ref{fig:advancing_bif_curve}) by its leading eigenmode, which is unstable. This case is more interesting, and we see that if $\Delta(0,\textbf{A}_2)<0$ (i.e. the perturbation `contracts' the \textbf{A}\textsubscript{2} interface) then the system returns to the stable state, whereas if $\Delta(0,\textbf{A}_2) > 0$ (i.e. a `stretch') then the contact line speed eventually diverges and the calculations fail to converge. Figure~\ref{fig:advancing_unstable_pert} shows the time-signal of $Y$ showing these two different dynamical outcomes. As shown by the inset labelled `Final State' the contact point velocity appears to diverge when $\theta_{\mathrm{app}}$ is close to 180$^\circ$. 

Similar outcomes occur for the receding contact-line problem. Perturbations, using the leading eigenmode, of the \textbf{R}\textsubscript{1} (figure~\ref{fig:receding_bif_curve}) stable state relax back to the stable equilibrium (results not shown). Figure~\ref{fig:receding_unstable_pert} shows the time evolution of perturbations from the \textbf{R}\textsubscript{2} (figure~\ref{fig:receding_bif_curve}) unstable state. Panel (a) shows the time-signal of $Y$ when $\Delta(0,\textbf{R}_2)<0$ and panel (b) when $\Delta(0,\textbf{R}_2)>0$. In the first case the system relaxes back to the \textbf{R}\textsubscript{1} state but if $\Delta(0,\textbf{R}_2)>0$ then, unlike the advancing contact line where the contact-line speed diverges, a thin-film develops that grows in size at a linear rate as $t\to\infty$ as shown in panel (b). 

These IVP calculations show that if we consider the class of perturbations representing `stretches' using the leading eigenmode only, then the indicator of whether the system returns to the stable state is that meniscus rise of the initial perturbation is smaller than the meniscus rise of the unstable state, i.e. the condition for stability is
\begin{align}
\Delta(0,\textbf{A}_2) &< 0,\qquad\mbox{Advancing Contact Line},\\
\Delta(0,\textbf{R}_2) &< 0,\qquad\mbox{Receding Contact Line}
\end{align}
In the language of dynamical systems the unstable state represents the `boundary of the basin of attraction' of the stable state, when considering simple stretches of the stable interface corresponding to perturbations using the leading eigenmode. The unstable state is therefore not just a trivial consequence of the steady bifurcation structure but also has an important role in the underlying transient dynamics. 

Another important observation is that the calculation in figure~\ref{fig:receding_unstable_pert}, for the receding contact line, provides evidence for the claim from \cite{snoeijer2012theory} that the solution curve represents the effective dynamics of the system ``in which the state of the solution moves quasi-statically along the solution curve''. In the receding case, and, although to a lesser degree, the advancing case, the trajectory of the system in the phase-space $(\overline{Ca},Y)$ is qualitatively similar to the steady solution curve when plotted in the same diagram. The inset diagrams in figures~\ref{fig:advancing_unstable_pert} and ~\ref{fig:receding_unstable_pert} labelled `Phase plane' show the steady solution curve and the trajectory of the system shown with an arrow. In both advancing and receding cases, when $\Delta(0,\textbf{A}_1/\textbf{R}_1)<0$ the trajectory follows a similar path to the solution curve. In the receding case, when $\Delta(0,\textbf{R}_2)>0$ the trajectory closely follows the upper reaches of the unstable branch. The similarity between the steady solution curve and the time-dependent trajectories will be discussed in more detail in the next section.

\begin{figure}
  \centering
\includegraphics[scale=0.26,trim= 0 0 0 0,clip]{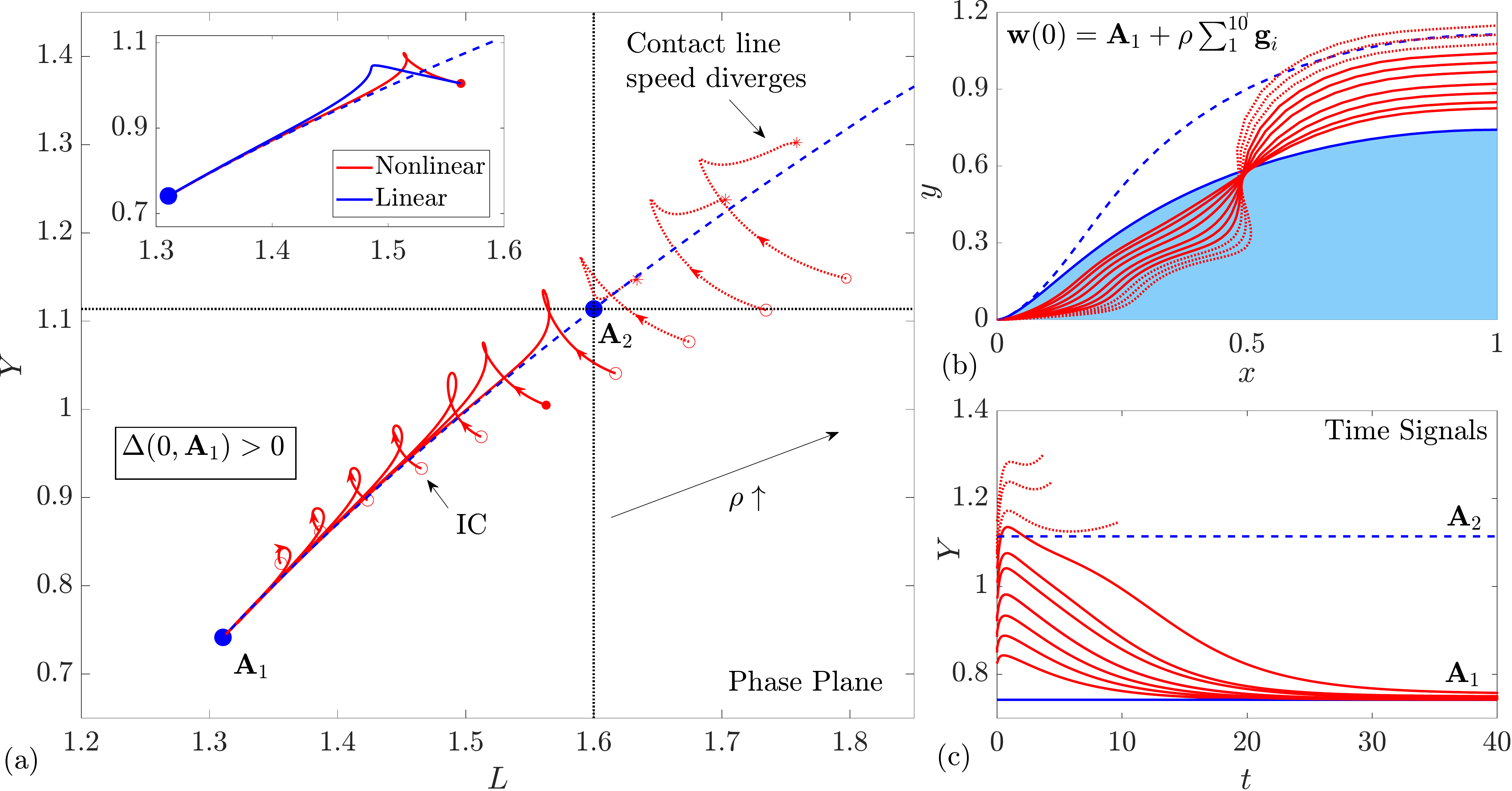}
	\caption{Time-dependent perturbations of the stable branch using the `corrugation' perturbations. This figure shows perturbations from the (stable) solution labelled \textbf{A}\textsubscript{1} in figure~\ref{fig:advancing_bif_curve} using \eqref{mixed_pert}. (a) Trajectories in the $(L,Y)$ phase plane projection. Initial perturbations are denoted by hollow circular markers, steady states are large solid circular markers and stars indicate that the contact line velocity has diverged. The solid trajectories correspond to initial perturbations that return to the steady state and dotted trajectories diverge. The dashed line indicates the unstable manifold of the \textbf{A}\textsubscript{2} state. The inset diagram shows the linear response given in \eqref{eigenmode_sum} for a particular initial condition marked with a solid circle in the main panel. (b) The initial conditions of the trajectories shown in (a) are shown as solid lines and the \textbf{A}\textsubscript{1} and \textbf{A}\textsubscript{2} steady states are shown using the shaded area and a dashed line, respectively. (c) The corresponding time signals of $Y$ for the trajectories shown in (a). Parameter values are $Ca = 0.934,\chi=0.1,\lambda = 0.1$.}
	\label{fig:mixed_perturbation_stable}
\end{figure}

\begin{figure}
  \centering
\includegraphics[scale=0.26]{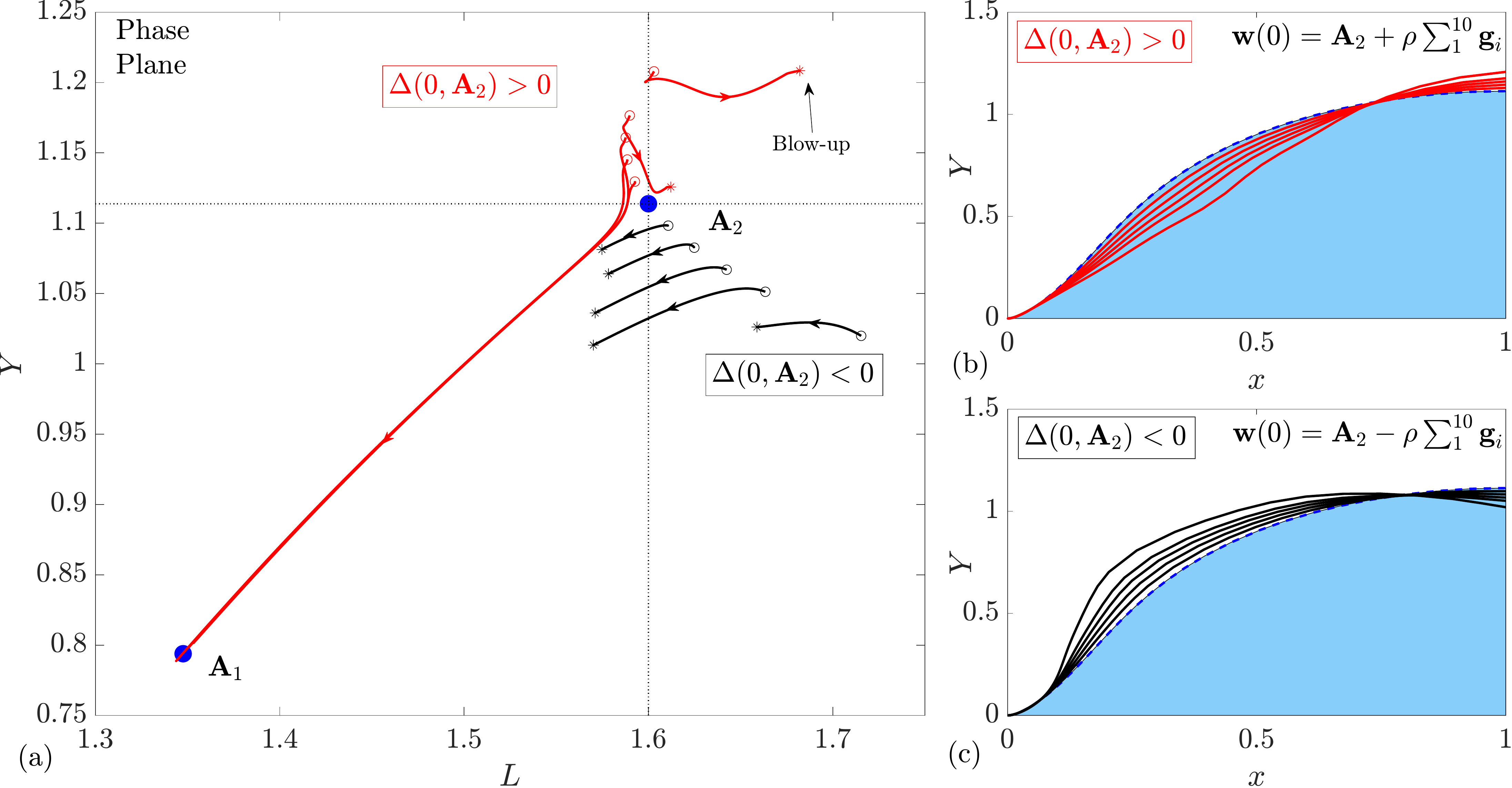}
	\caption{Time-dependent perturbations of the unstable branch. This figure shows perturbations from the (unstable) solution labelled \textbf{A}\textsubscript{2} in figure~\ref{fig:advancing_bif_curve} using \eqref{mixed_pert}. (a) Trajectories in the $(L,Y)$ phase plane projection. Initial perturbations are denoted by hollow circular markers, steady states are large solid circular markers and stars indicate that the contact-line velocity has diverged. (b) and (c) The initial interfaces compared to the stable and unstable steady states. Parameter values are $Ca = 0.934,\chi=0.1,\lambda = 0.1$.}
\label{fig:mixed_perturbation_unstable}
\end{figure}

\subsubsection{`Corrugation' perturbations}

We now concentrate on the `corrugation' perturbations which arise from taking more eigenmodes in the perturbation expansion (\eqref{mixed_stable_pert}). As discussed in the previous section this class of perturbation includes higher modes, which we might reasonably expect in a noisy physical system. We choose the number of eigenmodes as $N=10$ for computational efficiency. We set the initial condition of the system therefore as
\bea
\textbf{w}(t=0) = \textbf{w}_{\star} + \rho\sum_{i=1}^{N}\textbf{g}_i + \mbox{c.c}, \qquad N = 10.
\label{mixed_pert}
\eea

We concentrate solely on the advancing case and, as before, examine perturbations from the stable \textbf{A}\textsubscript{1} and unstable \textbf{A}\textsubscript{2} (figure~\ref{fig:advancing_bif_curve}) states using the IC prescribed in \eqref{mixed_pert}. Figure~\ref{fig:mixed_perturbation_stable} shows the time-dependent results for incrementally increasing values of $\rho$ in \eqref{mixed_pert}. Panel (a) shows trajectories in the $(L,Y)$ phase plane projection, where $L$ is the total arclength of the interface. To help orient the trajectories we have added artificial axes which are centred on the unstable \textbf{A}\textsubscript{2} state. 

After perturbing the \textbf{A}\textsubscript{1} state, the system experiences initial transient growth in the value of $Y$, corresponding to the weak attraction of the unstable \textbf{A}\textsubscript{2} state, before either the system settles to the stable state (solid trajectories) or the contact-line velocity diverges (dotted trajectories), indicated by the asterisks in panel (a). The dashed line between the \textbf{A}\textsubscript{1} and \textbf{A}\textsubscript{2} state is the response of the system if only the leading eigenmode is retained in the initial perturbation, as considered in the previous section on `stretch' modes which can be interpreted as the approximate unstable manifold of the \textbf{A}\textsubscript{2} state.

The nonlinear trajectories that return to the stable state all eventually collapse on the unstable manifold, once the higher-order modes have sufficiently decayed. The combination of higher-order modes in the initial perturbation does however cause initial transient growth in the system, despite the corresponding eigenvalues being highly stable as can be seen by the inset in the main diagram where the linear response, predicted by \eqref{eigenmode_sum}, of a particular trajectory (initial condition is marked with a solid circle) is compared to the full system response.

The unstable \textbf{A}\textsubscript{2} state, as indicated by a large circular symbol, plays a crucial role in the partition of behaviours. We make the important observation that initial conditions with an interface length, $L$, sufficiently larger than the unstable state will eventually become unstable. In contrast, there are also initial conditions with $\Delta(0,\textbf{A}_2)<0$ (see initial conditions in the lower-half plane of panel(a)) which still become unstable due to the initial transient growth caused by the combination of higher-order modes; in direct contradiction to the simpler perturbation of the `stretch' modes, where if $\Delta(0,\textbf{A}_2)<0$, then the system would relax to the stable state.

By examining the trajectories that explore the vicinity of the unstable branch in figure~\ref{fig:mixed_perturbation_stable}, it is not unreasonable to hypothesise that by continually refining the value of $\rho$ in the initial perturbation the system would be able to stay in the vicinity of the unstable state for an arbitrary time-period; where the dynamics of the system are dominated by the stable eigenmodes of the unstable branch. This behaviour is typically indicative of interpreting the unstable branch as an edge state, commonly used to describe weakly unstable states in the transition to turbulence and other fluid dynamics problems \citep[see][for a description of an edge state]{kerswell2014optimization}. In these scenarios the weakly unstable edge state acts as an `edge' between two dynamical outcomes; in our problem it separates the system returning to the stable state or the system diverging.

The role of the unstable state is further emphasised in figure~\ref{fig:mixed_perturbation_unstable} for perturbations from the unstable \textbf{A}\textsubscript{2} branch. In panel (a) different colour trajectories correspond to whether a positive or negative $\rho$ is chosen in the IC given by \eqref{mixed_pert}, and panels (b) and (c) show the initial conditions compared to the \textbf{A}\textsubscript{2} steady state. Again, the system outcomes are partitioned by the presence of the unstable state which `deflects' the system to either relax back to the stable state or the contact line velocity diverges. We note that in this case if $\Delta(0,\textbf{A}_2)<0$ (ICs in the lower half plane of panel (a)) then the system will become unstable which is again in direct contradiction to the `stretch' mode perturbations. These calculations demonstrate that $L$ is a more robust indicator of dynamical outcome; in this figure, all trajectories in the right side of the unstable state eventually experience instability contact line velocity divergence.

\subsubsection{Physical Significance of the perturbations}

We now relate these results to the physical system. Firstly, we make the key observation that the system is able to experience instability for $Ca<Ca_{\mathrm{crit}}$ due to finite-amplitude perturbations of the stable state. Furthermore, these results suggest that the system is more susceptible to instability caused by perturbations that increase the total arclength rather than increase the meniscus rise; corrugations, representing `noise', that increase the arclength, but not the meniscus rise, are more dangerous. Given that noise is a ubiquitous phenomena in physical systems we would expect to see this realised in a system with $Ca<Ca_{\mathrm{crit}}$. Secondly, the extent to which we are able to perturb the stable \textbf{A}\textsubscript{1}/\textbf{R}\textsubscript{1} state so that the system remains stable is dictated by the information encoded in the unstable \textbf{A}\textsubscript{2}/\textbf{R}\textsubscript{2} steady state, in particular whether the perturbation causes the interface length to increase beyond the length of the unstable steady state. A consequence of this is that the closer $Ca$ is to $Ca_{\mathrm{crit}}$ (from below), the smaller the finite amplitude is required, as the stable and unstable branches are increasingly approaching each other. 

As a conclusion, finite disturbances from the steady stable state potentially lead to instability before that interface might become unstable otherwise (due to a lack of steady state). Minimising ambient disturbances and minimising fluctuations in $Ca$ would help maintain a stable interface for higher $Ca$. The physical mechanisms underlying the observations about stability/instability are not straightforward to address due to the highly nonlinear nature of the problem.  Moreover, three-dimensional perturbations will likely be important in practice and introduce additional physical mechanisms.  Such 3D calculations are exceedingly challenging and are currently the focus of ongoing research, so we defer a detailed study of physical mechanisms for future work.

\subsection{Dynamics `beyond' the fold: $Ca>Ca_{\mathrm{crit}}$}  
\begin{figure}
  \centering
\includegraphics[scale=0.26]{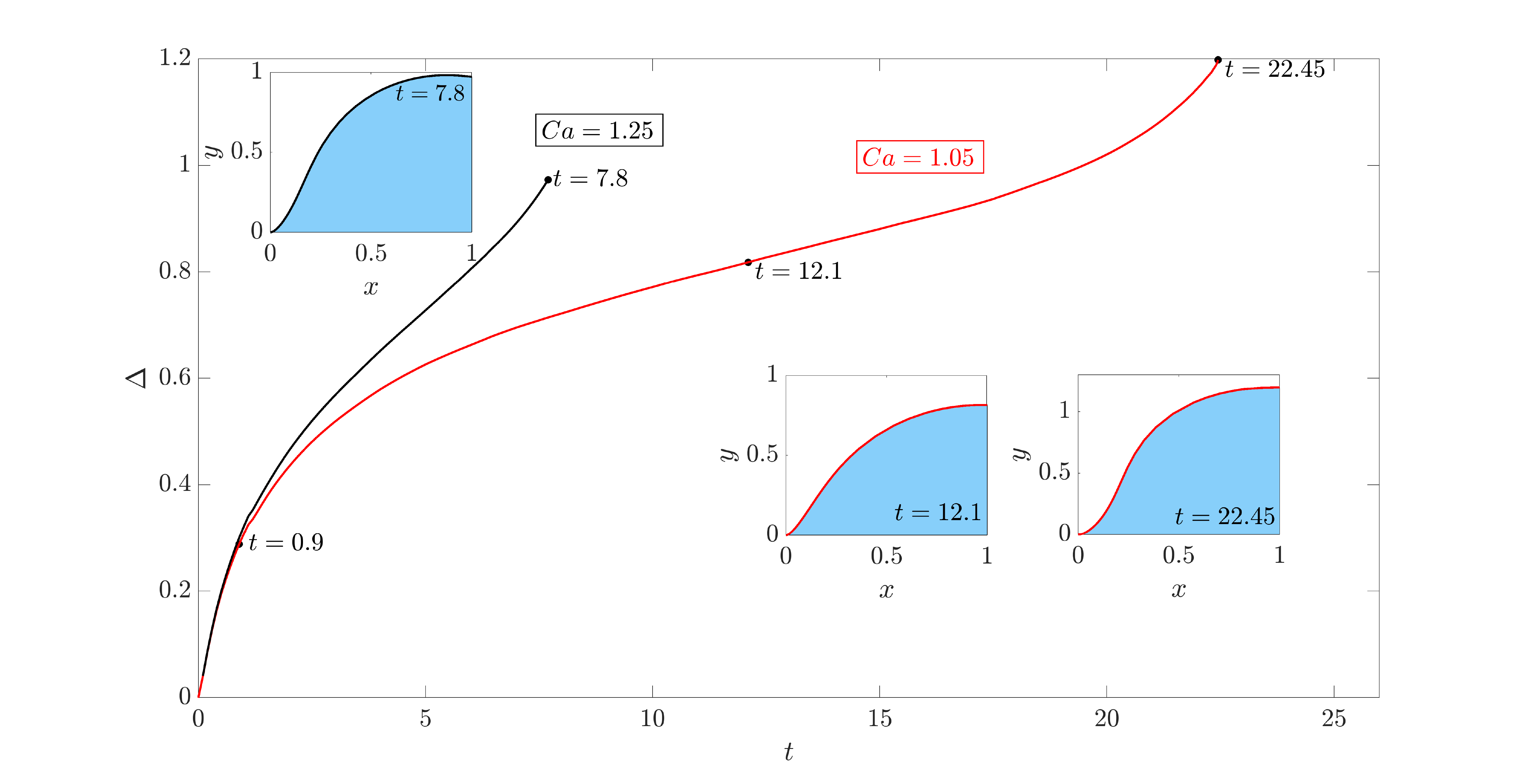}
\caption{Time-dependent evolution for a advancing contact line when $Ca>Ca_{\mathrm{crit}}$ and the system is at rest initially. The different colour curves indicate different values of $Ca = 1.05,1.25$. The inset panels on the left are the interface profiles corresponding to the markers labelled in the main figure. Other parameter values are $\chi=0.1,\lambda = 0.1$.}
\label{fig:advancing_unstable_IVP}
\end{figure}

\begin{figure}
  \centering
\includegraphics[scale=0.26,trim=0 0 0 0,clip]{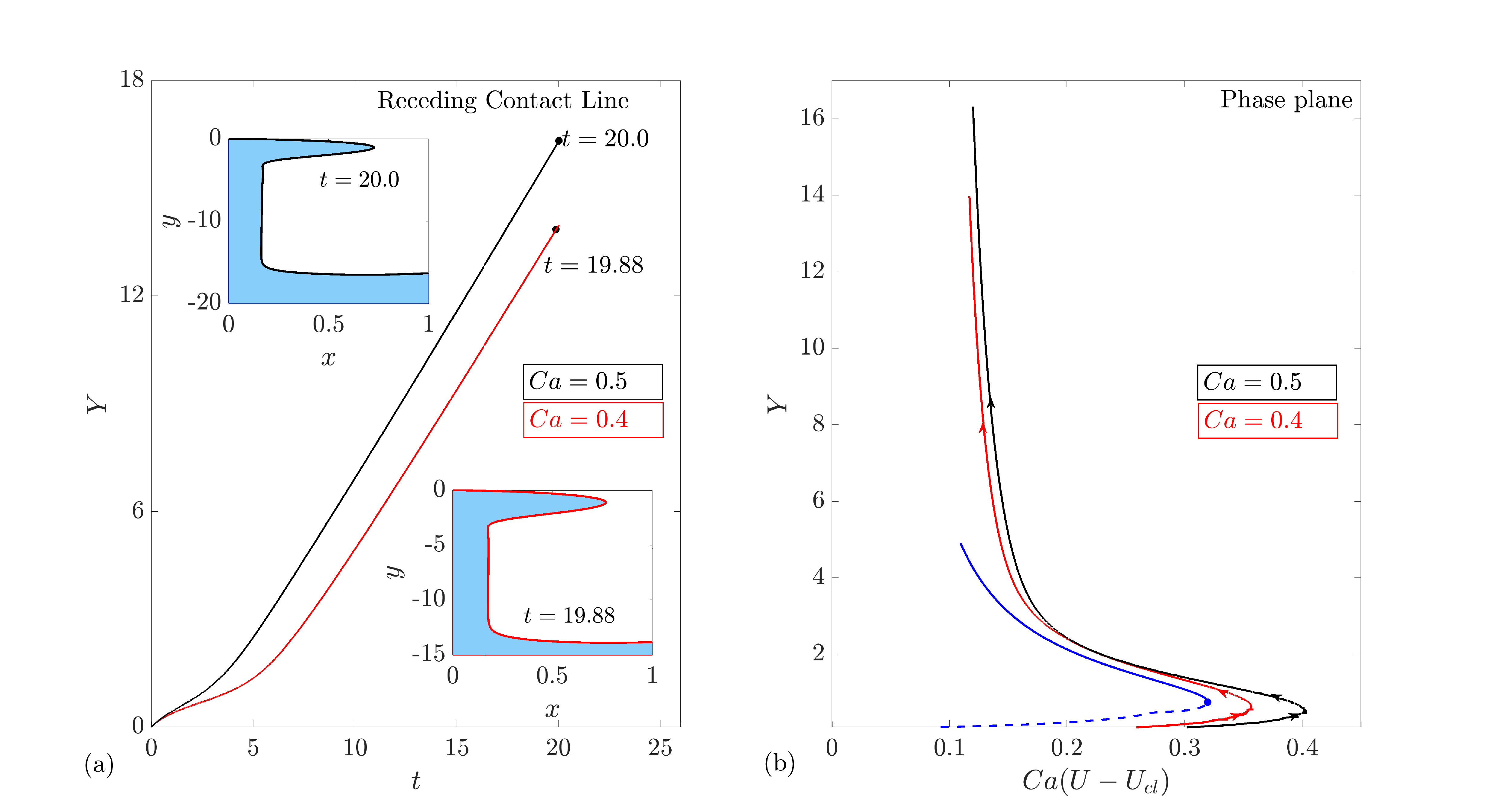}
\caption{Time-dependent evolution for a receding contact line when $Ca>Ca_{\mathrm{crit}}$ and the system is at rest initially for $\lambda = 0.1,\chi=0$. The different colour curves indicate different values of $Ca = 0.4,0.5$. (a) The time-signal of $Y$ with the final interface profiles in the inset. (b) The trajectories (curves with arrows) in the $(\overline{Ca},Y)$ phase-plane projection compared with the steady bifurcation curve (curve without arrows). Other parameter values are $\chi=0,\lambda = 0.1$.}
\label{fig:receding_unstable_IVP}
\end{figure}
\begin{figure}
  \centering
  \includegraphics[scale=0.26]{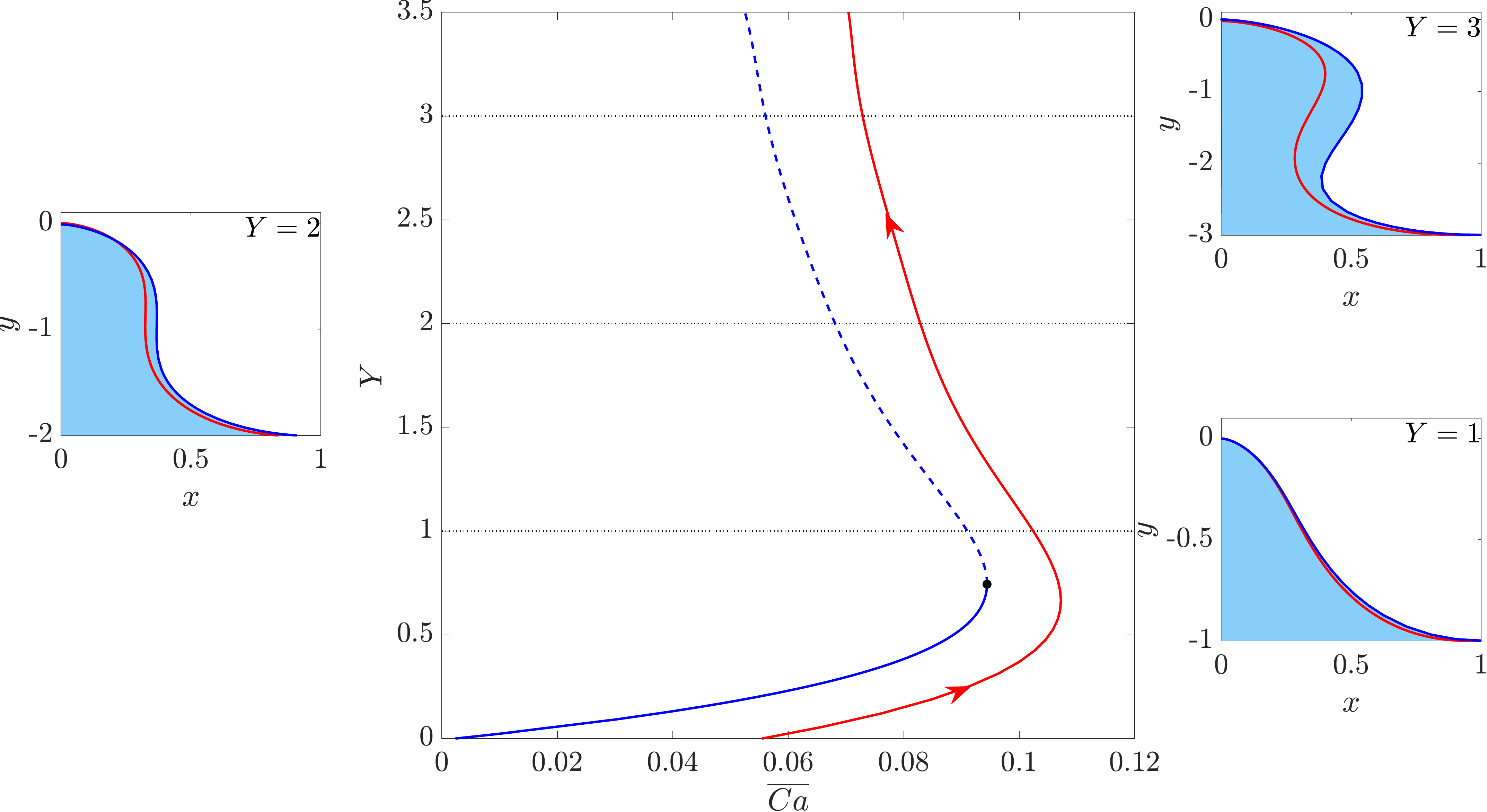}
  \caption{Time-dependent evolution for a receding contact line when $Ca>Ca_{\mathrm{crit}}$ and the system is at rest initially. The curve with arrows is the time-dependent trajectory whilst the curve without arrows is the steady solution curve in the $(\overline{Ca},Y)$ projection. The panels on the outside of the main panel compare the interface profiles for the time-dependent evolution and the steady solution branches at the same values of $Y = 1,2,3$. Parameter values are $\lambda= 0.01,\chi=0.1,Ca = 0.15$.}
  \label{fig:receding_comparison}
\end{figure}

We now turn our attention to starting the system from rest with a flat interface beyond the fold, i.e. $Ca>Ca_{\mathrm{crit}}$, so that we are able to answer the second question stated at the start of \S~\ref{sec:transient_dynamics}. The IC is
\bea
\textbf{w}(t=0) = [\textbf{u}(0),p(0),p_1(0),\textbf{r}(0)]^T = [\textbf{0},\textbf{0},\textbf{0},(x,0)]^T.
\label{IC_flat}
\eea
For values of $Ca$ exceeding the critical value there are no (known) steady states which can influence the system, unlike in the previous section. For the advancing case, the system quickly diverges with the contact line speed diverging rapidly causing the calculations to fail to converge, as shown in figure~\ref{fig:advancing_unstable_IVP} where the time-signal of $Y$ is measured along with time snapshots of the interface at the times indicated for $Ca = 1.05,1.25$. We emphasise that the system is not attracted to a different steady state of the system that may exist and we are unable to find any additional steady states beyond the fold bifurcation. 

Prior to the calculations ceasing to converge, the pressure gradients are particularly strong near the contact point and as explained earlier we would expect the fluid pressure gradient to exceed the capillary stress gradient so that the fluid phase is unable to pump fluid out near the contact line.

For the receding case the contact line velocity does not diverge and a thin-film develops. Figure~\ref{fig:receding_unstable_IVP} (a) shows the time-signal of $Y$ in the receding case when $Ca = 0.4,0.5 > Ca_{\mathrm{crit}}$. The system, although not approaching a steady state, forms a coherent structure whose meniscus rise grows at a constant velocity, similar to the structure seen in figure~\ref{fig:receding_unstable_pert}(a). By examining the phase space, shown in figure~\ref{fig:receding_unstable_IVP}(b), we can see that the system trajectory closely matches the steady solution curve and approaches a fixed value of $\overline{Ca}$ as $t\to\infty$ which appears to be independent of $Ca$, indicating the contact-line region and the thin-film region are essentially de-coupled by this point.

Unlike the advancing case, the time-dependent trajectories and the steady solution curve for the receding problem, when plotted in $(\overline{Ca},Y)$, are qualitatively similar. Figure~\ref{fig:receding_comparison} shows the comparison of the time-dependent trajectory with the steady solution curve for $\lambda = 0.01,\chi = 0.1,Ca = 0.15>Ca_{\mathrm{crit}}$ as well as comparing the actual interface profiles for the time-dependent calculation and the corresponding steady solution for the same value of $Y$ (inset figures). It is striking how close the two corresponding interface profiles are, although we note that in the time-dependent case the horizontal height of the inflection point is smaller than that of the steady solution in all of the cases. This again shows compelling evidence that in the receding contact line problem the unsteady solution branch represents the effective dynamics of the system.

\section{Discussion}
\label{sec:discuss}

We have developed a time-dependent hybrid model and utilised ideas from dynamical systems theory to investigate the advancing and receding contact-line problem. By solving a generalised eigenproblem numerically and performing IVP simulations we have demonstrated that far from being passive, the unstable branch of the bifurcation structure plays a subtle role in the underlying time-dependent behaviour, from a dynamical systems perspective.

By perturbing the stable branch using the eigenmodes we have demonstrated that the unstable branch represents the basin boundary of attraction of the stable solution and it has a profound effect on the eventual evolution of the system. We have demonstrated that perturbations that cause the interface to stretch are robust, in that provided the stretch does not exceed that of the unstable branch the system returns to the stable state. In contrast, perturbations that increase the overall length of the interface by adding `corrugations' are more dangerous and the system instability is more susceptible to this class of perturbations. This information may be helpful in physical systems as a means of flow control as knowing the structure of the unstable eigenmodes may enable us to stabilise the system using suction/injection techniques. We note that the perturbations we have considered here are purely theoretical eigenmode perturbations and it would be of interest to examine the stability of the contact-line when physically perturbed by, for example, the effect of surface defects on the moving substrate. This is part of the authors' current research. 

In addition, by performing time-dependent calculations we have shown that trajectories in phase space qualitatively match the steady bifurcation structure. In both the receding and advancing cases, the solution curve describes what \cite{snoeijer2012theory} termed the `effective dynamics' of the system. The bifurcation structure remains structurally stable (i.e. the stable-fold-unstable branch structure does not change) as $\chi$ and $\lambda$ is varied, (see figure~\ref{fig:ca_crit_evolution}) and hence we predict the overall qualitative behaviour to be similar, i.e. the unstable branch is generically the basin boundary of attraction for systems of this nature.

Gravity and inertial effects can easily be added to the model and it is interesting what effect these would have on the bifurcation structure and stability analysis. We have already performed some preliminary calculations incorporating gravity and have found that the bifurcation structure experiences the same multiple saddle-node bifurcations as predicted by the lubrication model \citep{snoeijer2012theory}. Our preliminary analysis shows that the entire upper branch past the first saddle-node bifurcation is unstable, but we have not pursued this in detail and leave this for future research. For inertial effects, we could expect other types of bifurcations, including Hopf bifurcations, which would introduce complex eigenvalues to the spectrum of $\sigma$ and lead to more complex transient behaviour. We also leave this avenue for future research.  

In our model we choose the simplest approach and set the dynamic contact angle to be constant. It is hotly debated whether this is indeed physically realistic and whether molecular or hydrodynamic effects are stronger near the contact line. A natural extension to the model would be to investigate the effect of having a contact angle that is dependent on the velocity of the plate, such as that proposed by molecular kinetic theory (MKT), \citep[see, for example][]{toledano2021closer,Blake2006} or the interface formation model \citep[see][]{shikhmurzaev2007capillary}. Provided the bifurcation structure remains intact (i.e. stable-fold-unstable) then we predict the dynamics will be qualitatively the same. This is the subject of a submitted article where we apply the hybrid model and stability algorithm developed here to account for a $Ca$-dependent contact angle observed in molecular simulations \citep{keeler2021micro}.

For the receding contact line, the instability that occurs manifests itself as the formation of a thin film and a capillary ridge, but for the advancing case the calculations are significantly more difficult to get a converged solution and we are only able to advance in time until air entrainment first starts to occur. The instability in this study is taken in a two-dimensional context but it is well-known that the `saw-tooth' patterns that emerge in an unstable advancing contact line are intrinsically three-dimensional. A natural extension to our model would be to consider transverse perturbations of the advancing contact line in the third dimension and perform a stability analysis, similar to our approach here. This has been done before in the receding case \citep{snoeijer2007part1} and the advancing case \citep{vandrethesis} using a lubrication model, but the time-dependent hybrid model developed here is a perfect testing ground for a three-dimensional calculation as the computational cost is relatively small compared to the full model, and we are also able to fully account for the velocity in the liquid, rather than use a lubrication model. This direction of research is currently being developed.

\section{Acknowledgments}

We acknowledge funding from EPSRC grants EP/N016602/1, EP/P020887/1, EP/S029966/1 and EP/P031684/1 and useful discussions with Terry Blake.

\appendix

\section{Derivation of Hybrid Model}\label{app:hybrid}

As the flow is approximately two-dimensional, the momentum equations in the fluid phase are
\bea
\pdiff{p_1}{x} = 0,\qquad \pdiff{p_1}{y} = \chi \pdiffn{v_1}{x}{2},
\eea
so the pressure in the fluid phase is a function of $y$ only and a straightforward two-fold integration of the momentum equation in \eqref{fluid_phase} yields an expression for $v_1$, i.e.
\bea
v_1 = \frac{1}{\chi}\left(\frac{1}{2}\pdiff{p_1}{y}x^2 + Ax + B\right),
\label{hybrid_velocity}
\eea
where $A$ and $B$ are constants of integration. We impose the Navier-slip condition on $x=0$ and impose continuity of velocity at the interface boundary, denoted by $x=h$, as in figure~\ref{fig:domain}(c). These conditions determine $A$ and $B$:
\bea
A = \frac{B - U\chi}{\lambda},\,\qquad B = \frac{\chi(h + \lambda v) - \frac{1}{2}\lambda (\partial p_{1}/\partial y) h^2}{\lambda + h},
\eea
where $v$ is the horizontal velocity of the liquid phase and $p_1$ is the fluid pressure evaluated at the interface. We have now determined the horizontal velocity in the fluid phase in terms of the (as yet) unknown pressure gradient, $\partial p_{1}/\partial y$ on the interface {(the vertical fluid velocity, $v_1$, can be recovered using \eqref{hybrid_velocity})}.

We now use conservation of mass to form an equation that determines the pressure in the fluid at the interface. The conservation equation, \eqref{fluid_phase}, can be written as
\bea
\textbf{u}_1\cdot\textbf{n} + \frac{1}{\chi}\pdiff{}{y}\left(\frac{1}{6}\pdiff{p_1}{y}h^3 + \frac{1}{2}Ah^2 + Bh\right) = 0.
\label{hybrid_equation2}
\eea
As $\hat{H}/\hat{L} \ll 1$ ($\hat{H},\hat{L}$ are typical horizontal and vertical length scales, respectively) then the arclength along the interface, $s$, measured from the contact point (see figure~\ref{fig:domain}(e)) is $s = y + O(\hat{H}/\hat{L})$. Therefore we can replace derivatives w.r.t $y$ with derivatives w.r.t $s$. This is preferable because it allows the interface to become multi-valued in numerical calculations. This means that $\textbf{u}_1\cdot\textbf{n} = u_1 + O(\hat{H}/\hat{L})$. The kinematic condition, \eqref{kinematic_condition}, therefore means $u = \partial h/\partial t + O(\hat{H}/\hat{L})$ and hence equation~\eqref{hybrid_equation2} becomes
\bea
\pdiff{h}{t} + \frac{1}{\chi}\pdiff{Q_1}{s},\qquad Q_1 = \frac{1}{6}\pdiff{p_1}{s}h^3 + \frac{1}{2}Ah^2 + Bh = 0,
\label{hybrid_equation1}
\eea
as stated in the main text. This equation determines the evolution of the pressure in the fluid phase on the interface. We set $p_1(s=L) = 0$ and $Q_1(s=0) = 0$ to ensure pressure is in equilibrium in at the right plate and there is no mass flux through the contact line. The velocity in the fluid phase can be recovered using \eqref{hybrid_velocity}. In this formulation the equations in the liquid phase are coupled to \eqref{hybrid_equation1} due to the presence of $v$ in the definition of the constant $A$. Furthermore, we couple the pressure and velocity in the fluid phase to the liquid phase through the dynamic boundary condition, i.e.
\bea
\boldsymbol{\tau}_1\cdot{\textbf{n}} = p_1\textbf{n}-\chi\pdiff{v_1}{x}\textbf{t},\qquad \textbf{x}\in\Gamma_4
\label{coupling}
\eea
which is standard when using a thin-film approximation \citep{oron1997long}.

\begin{figure}
  \centering
  \includegraphics[scale=0.3]{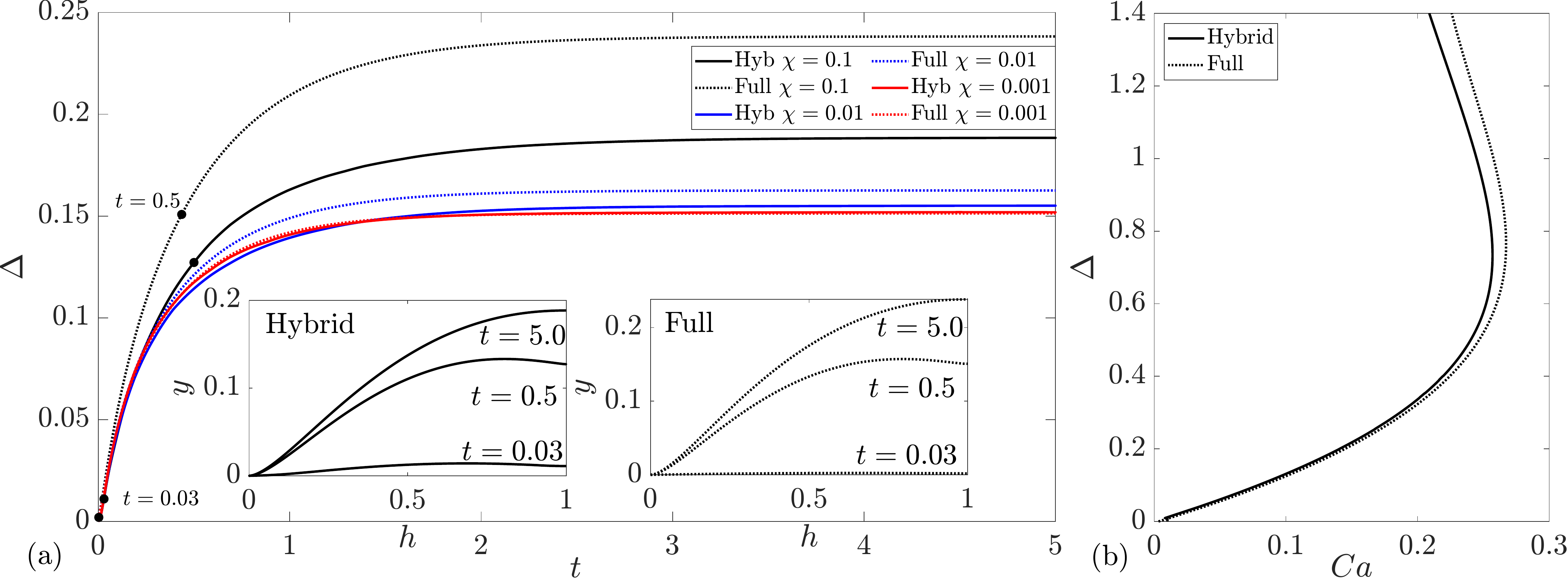}
  \caption{Comparison of the hybrid and full model for the advancing contact-line. (a) Time signals of the meniscus rise, $Y = |y(1) - y(0)|$, of the hybrid (solid) and full (dashed) models when $\lambda = 0.1,Ca = 0.2$. The different colours indicate $\chi=0.1,0.01,0.001$. As the main panel shows the system relaxes to a stable state and as $\chi\to 0$ the two models converge. The inset panels show interface profiles at sampled times indicated by the labels. (b) The comparison of the steady solution curve when $\lambda = 0.1,\chi = 0.1$ for the hybrid (solid curve) and full (dotted curve) models. The curves are calculated using the numerical method described in~\S\ref{sec:num_method}}
  \label{fig:validation}
\end{figure}

{To validate the hybrid model we performed a series of time-dependent IVP calculations starting the system from rest ($\textbf{u}_i = \textbf{0}$) with a flat interface. If we choose $Ca<Ca_{\mathrm{Crit}}$ we expect the system to eventually relax to a stable state. As $\chi\to 0$ the full model and hybrid model should converge.} Panel (a) of figure~\ref{fig:validation} shows the time-signal of the meniscus rise, $Y$, for different viscosity ratios (different colours) for the full model (dotted lines) and the hybrid model (solid lines). In all cases the interface eventually relaxes to a stable state, as shown by the time signals in the main figure. It is also clear that for the smallest value of $\chi$ in the figure the full and hybrid model are virtually indistinguishable. This is because for small $\chi$, the fluid only has an influence on the liquid in thin films in front of the contact line, and this is where the lubrication model is valid. In contrast, at moderate $\chi$ the influence of the fluid is felt everywhere, the films are thicker/non-existent and this approximation loses accuracy. In other words, this approximation works best when the fluid is a gas.

Panel (b) of figure~\ref{fig:validation} shows the comparison of the solution curve in the receding problem for the full and hybrid model. As can be seen the value of $Ca_{\mathrm{Crit}}$ changes only minimally for the hybrid model when compared to the full model.

\section{Validation of eigenvalue calculation}\label{sec:eigenmode_validation}

\begin{figure}
\centering
  \includegraphics[scale=0.3]{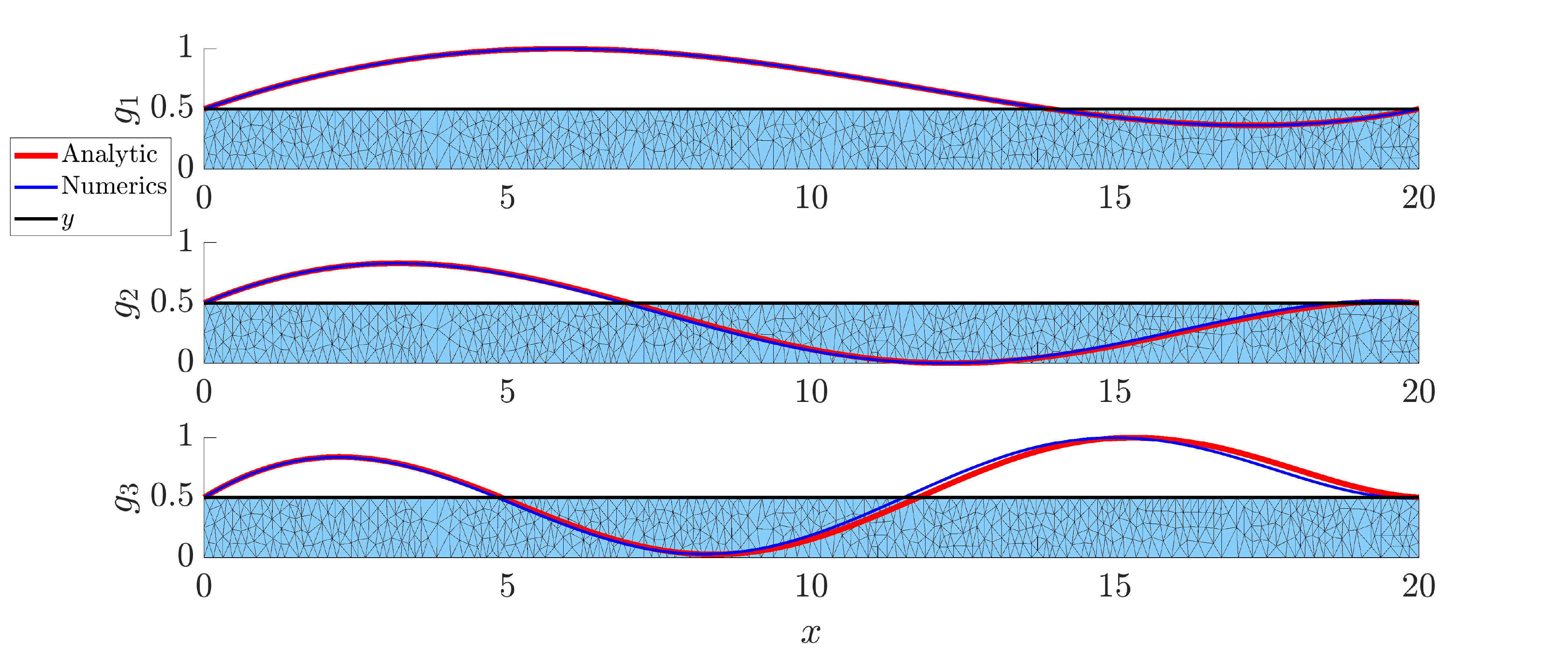}
  \caption{Comparison of eigenmodes, $H/L = 40$, $C = 3.0$. The different colour profiles indicate the analytic eigenmodes from \eqref{analytic_eigenmode} and those obtained from the numerical calculations. The top figure is the eigenmode $g_1$, corresponding to the largest eigenvalue whilst the lower panels show $g_2$ and $g_3$, the eigenmodes for the next two largest eigenvalues.}
  \label{fig:eigenmode_validation}
\end{figure}

We can validate the solutions of \eqref{eigen_eqn} against a situation where analytic eigenmodes are known. If we assume the wall is static ($U=0$) and that $\hat{H}/\hat{L}\gg 1$, corresponding to a short fat pool of liquid at the bottom of the channel with no-slip beneath it, then we can also apply a lubrication approximation to the liquid domain. If the fluid is treated as a vacuum we have the following equation for the vertical height of the interface:
\bea
y_t + C\left(y^3y_{xxx}\right)_x = 0,
\label{lub_eigen_eqn}
\eea
where $C$ is a known constant containing non-dimensional system parameters (in this case we choose $C=3$). We choose the boundary conditions $y(0) = y(L) = y_0, y_{xxx}(0) = y_{xxx}(L) = 0$ which describe a pinned contact line and no flux through the walls so that the resulting interface is flat in equilibrium. In this case the set of unknowns is `one-dimensional' in that $w = [y]$ and using the perturbation in \eqref{perturbation} yields a set of equations where analytic progress can be made. It can be shown that the eigenvalues and eigenmodes of \eqref{lub_eigen_eqn} can be written as
\begin{multline}
  g_n = a_n\left[\frac{\sinh(\sigma_n^{1/4}L) + \sin(\sigma_n^{1/4}L)}{\cos(\sigma_n^{1/4}L) - \cosh(\sigma_n^{1/4}L)}(\cosh(\sigma_n^{1/4}x) - \cos(\sigma_n^{1/4}x)) + \right.\\
    \left. \sinh(\sigma_n^{1/4}x) + \sin(\sigma_n^{1/4}x)\right],\qquad \sigma_n \approx \left(\frac{\pi/2 + n\pi}{L}\right)^4. 
\label{analytic_eigenmode}
\end{multline}
These non-trivial analytic expressions can be used to validate our numerical stability calculations. We chose a computational domain with $H/L = 40$, to reflect $\hat{H}/\hat{L}\gg 1$, and calculate the corresponding eigenmodes numerically using the hybrid model described in the previous section. Figure~\ref{fig:eigenmode_validation} shows the first three eigenmodes corresponding to the largest three eigenmodes, $n=1,2,3$ as calculated by the hybrid model and the analytic solution (different colour curves). These solutions are stable as the eigenspectra lies exclusively in the left-hand complex plane and the agreement between the numerical calculations and the analytic solution obtained from the lubrication model is excellent, giving us confidence in our computational framework.

\vskip2pc

\bibliographystyle{RS} 

\begin{thebibliography}{57}
\expandafter\ifx\csname natexlab\endcsname\relax\def\natexlab#1{#1}\fi
\def\au#1{#1} \def\ed#1{#1} \def\yr#1{#1}\def\at#1{#1}\def\jt#1{\textit{#1}}
  \def\bt#1{#1}\def\bvol#1{\textbf{#1}} \def\vol#1{#1} \def\pg#1{#1}
  \def\publ#1{#1}\def\arxiv#1{#1}\def\org#1{#1}\def\st#1{\textit{#1}}

\bibitem[Afkhami {\em et~al.\/}(2020)Afkhami, Gambaryan-Roisman \&
  Pismen]{afkhami2020challenges}
{\sc \au{Afkhami, S.}, \au{Gambaryan-Roisman, T.} \& \au{Pismen, L.~M.}}
  \yr{2020}  \at{Challenges in nanoscale physics of wetting phenomomen}.
  \jt{Euro. Phys. J. Spec. Top.}  \bvol{229}~(10),  \pg{1735--1738}.

\bibitem[Blake(2006)]{Blake2006}
{\sc \au{Blake, T.~D.}} \yr{2006}  \at{The physics of moving wetting lines}.
  \jt{J. Coll. Inter. Sci.}  \bvol{299},  \pg{1–13}.

\bibitem[Bonn {\em et~al.\/}(2009)Bonn, Eggers, Indekeu, Meunier \&
  Rolley]{bonn2009wetting}
{\sc \au{Bonn, D.}, \au{Eggers, J.}, \au{Indekeu, J.}, \au{Meunier, J.} \&
  \au{Rolley, E.}} \yr{2009}  \at{Wetting and spreading}.  \jt{Rev. Mod. Phys.}
   \bvol{81}~(2),  \pg{739}.

\bibitem[Chan {\em et~al.\/}(2020)Chan, Kamal, Snoeijer, Sprittles \&
  Eggers]{chan2020cox}
{\sc \au{Chan, T.~S.}, \au{Kamal, C.}, \au{Snoeijer, J.~H.}, \au{Sprittles,
  J.~E.} \& \au{Eggers, J.}} \yr{2020}  \at{Cox--voinov theory with slip}.
  \jt{J. Fluid Mech.}  \bvol{900},  \pg{A8}.

\bibitem[Chan {\em et~al.\/}(2012)Chan, Snoeijer \& Eggers]{snoeijer2012theory}
{\sc \au{Chan, T.~S.}, \au{Snoeijer, J.~H.} \& \au{Eggers, J.}} \yr{2012}
  \at{Theory of the forced wetting transition}.  \jt{Phys. Fluids}
  \bvol{24}~(072104).

\bibitem[Chan {\em et~al.\/}(2013)Chan, Srivastava, Marchand, Andreotti,
  Biferale, Toschi \& Snoeijer]{chan2013hydrodynamics}
{\sc \au{Chan, T.~S.}, \au{Srivastava, S.}, \au{Marchand, A.}, \au{Andreotti,
  B.}, \au{Biferale, L.}, \au{Toschi, F.} \& \au{Snoeijer, J.~H.}} \yr{2013}
  \at{Hydrodynamics of air entrainment by moving contact lines}.  \jt{Phys.
  Fluids}  \bvol{25}~(7),  \pg{074105}.

\bibitem[Charitatos {\em et~al.\/}(2020)Charitatos, Suszynski, Carvalho \&
  Kumar]{charitatos2020dynamic}
{\sc \au{Charitatos, V.}, \au{Suszynski, W.~J.}, \au{Carvalho, M.~S.} \&
  \au{Kumar, S.}} \yr{2020}  \at{Dynamic wetting failure in shear-thinning and
  shear-thickening liquids}.  \jt{J. Fluid Mech.}  \bvol{892}.

\bibitem[Christodoulou \& Scriven(1988)]{christodoulou1988finding}
{\sc \au{Christodoulou, K.~N.} \& \au{Scriven, L.~E.}} \yr{1988}  \at{Finding
  leading modes of a viscous free surface flow: An asymmetric generalized
  eigenproblem}.  \jt{J. Sci. Comp.}  \bvol{3}~(4),  \pg{355--406}.

\bibitem[Cox(1986)]{cox1985part1}
{\sc \au{Cox, R.~G.}} \yr{1986}  \at{The dynamics of the spreading of liquids
  on a solid surface. part 1. viscous flow}.  \jt{J. Fluid Mech.}  \bvol{168},
  \pg{169--194}.

\bibitem[Doedel(2007)]{doedel}
{\sc \au{Doedel, Eusebius~J}} \yr{2007}  \at{Lecture notes on numerical
  analysis of nonlinear equations}.  \bt{In {\em Numerical Continuation Methods
  for dynamical systems\/}},  \pg{pp. 1--49}.  \publ{Springer}.

\bibitem[Eckhardt {\em et~al.\/}(2008)Eckhardt, Faisst, Schmiegel \&
  Schneider]{eckhardt2008dynamical}
{\sc \au{Eckhardt, B.}, \au{Faisst, H.}, \au{Schmiegel, A.} \& \au{Schneider,
  T.~M.}} \yr{2008}  \at{Dynamical systems and the transition to turbulence in
  linearly stable shear flows}.  \jt{Phil. Trans. R. Soc. Lond. A}
  \bvol{366}~(1868),  \pg{1297--1315}.

\bibitem[Eggers(2004{\natexlab{{\em a\/}}})]{eggers2004forced}
{\sc \au{Eggers, J.}} \yr{2004{\natexlab{{\em a\/}}}}  \at{Hydrodynamic theory
  of forced dewetting}.  \jt{Phys. Rev. Letters}  \bvol{96}~(174504).

\bibitem[Eggers(2004{\natexlab{{\em b\/}}})]{eggers2004towards}
{\sc \au{Eggers, J.}} \yr{2004{\natexlab{{\em b\/}}}}  \at{Towards a
  description of contact line motion at higher capillary numbers}.  \jt{Phys.
  Fluids.}  \bvol{16}~(9),  \pg{3491 -- 3494}.

\bibitem[Eggers(2005)]{eggers2005existence}
{\sc \au{Eggers, J.}} \yr{2005}  \at{Existence of receding and advancing
  contact lines}.  \jt{Phys. Fluids}  \bvol{17}~(082106).

\bibitem[Fern{\'a}ndez-Toledano {\em et~al.\/}(2021)Fern{\'a}ndez-Toledano,
  Blake \& Coninck]{toledano2021closer}
{\sc \au{Fern{\'a}ndez-Toledano, J.~C.}, \au{Blake, T.~D.} \& \au{Coninck,
  J.~D.}} \yr{2021}  \at{Taking a closer look: A molecular-dynamics
  investigation of microscopic and apparent dynamic contact angles}.  \jt{J.
  Coll. Inter. Sci.}  \bvol{587},  \pg{311--323}.

\bibitem[Gaillard {\em et~al.\/}(2020)Gaillard, Keeler, Thompson, Hazel \&
  Juel]{gaillard2020life}
{\sc \au{Gaillard, A.}, \au{Keeler, J.~S.}, \au{Thompson, A, J.}, \au{Hazel,
  A.~H.} \& \au{Juel, A.}} \yr{2020}  \at{Life and fate of a bubble in a
  constricted hele-shaw channel}.  \jt{J. Fluid. Mech}  \bvol{122},
  \pg{12--12}.

\bibitem[Gallino {\em et~al.\/}(2018)Gallino, Schneider \&
  Gallaire]{gallino2018edge}
{\sc \au{Gallino, G.}, \au{Schneider, T.~M.} \& \au{Gallaire, F.}} \yr{2018}
  \at{Edge states control droplet breakup in subcritical extensional flows}.
  \jt{Phys. Rev. Fluids}  \bvol{3},  \pg{073603}.

\bibitem[He(2020)]{he2020long}
{\sc \au{He, M.}} \yr{2020}  \at{Long-time evolution of interfacial structure
  of partial wetting}.  \jt{Phys. Rev. Fluids}  \bvol{5}~(11),  \pg{114001}.

\bibitem[He \& Nagel(2019)]{he2019characteristic}
{\sc \au{He, M.} \& \au{Nagel, S.~R.}} \yr{2019}  \at{Characteristic
  interfacial structure behind a rapidly moving contact line}.  \jt{Phys. Rev.
  Lett.}  \bvol{122}~(1),  \pg{018001}.

\bibitem[Heil \& Hazel(2006)]{heil2006oomph}
{\sc \au{Heil, M.} \& \au{Hazel, A.~L.}} \yr{2006}  \at{oomph-lib--an
  object-oriented multi-physics finite-element library}.  \bt{In {\em
  Fluid-structure interaction\/}},  \pg{pp. 19--49}.  \publ{Springer}.

\bibitem[Heroux {\em et~al.\/}(2003)Heroux, Bartlett, Howle, Hoekstra, Hu,
  Kolda, Lehoucq, Long, Pawlowski, Phipps, Salinger, Thornquist, Tuminaro,
  Willenbring \& Williams]{herouxtrilnos}
{\sc \au{Heroux, M.}, \au{Bartlett, R.}, \au{Howle, V.}, \au{Hoekstra, R.},
  \au{Hu, J.}, \au{Kolda, T.}, \au{Lehoucq, R.}, \au{Long, K.}, \au{Pawlowski,
  R.}, \au{Phipps, E.}, \au{Salinger, A.}, \au{Thornquist, H.}, \au{Tuminaro,
  R.}, \au{Willenbring, J.} \& \au{Williams, A.}} \yr{2003}  \bt{An overview of
  trilinos}. {\em Tech. Rep.\/}.  \org{SAND2003-2927. Sandia National
  Laboratories}.

\bibitem[Huh \& Scriven(1971)]{huh1971hydrodynamic}
{\sc \au{Huh, C.} \& \au{Scriven, L.~E.}} \yr{1971}  \at{Hydrodynamic model of
  steady movement of a solid/liquid/fluid contact line}.  \jt{J. Coll. Inter.
  Sci.}  \bvol{35}~(1),  \pg{85--101}.

\bibitem[Jacqmin(2004)]{jacqmin2004onset}
{\sc \au{Jacqmin, D.}} \yr{2004}  \at{Onset of wetting failure in
  liquid–liquid systems}.  \jt{J. Fluid Mech.}  \bvol{517}.

\bibitem[Kamal {\em et~al.\/}(2019)Kamal, Sprittles, Snoeijer \&
  Eggers]{kamal2019dynamic}
{\sc \au{Kamal, C.}, \au{Sprittles, J.~E.}, \au{Snoeijer, J.~H.} \& \au{Eggers,
  J.}} \yr{2019}  \at{Dynamic drying transition via free-surface cusps}.
  \jt{J. Fluid Mech.}  \bvol{858}.

\bibitem[Keeler {\em et~al.\/}(2021)Keeler, Blake, Lockerby \&
  Sprittles]{keeler2021micro}
{\sc \au{Keeler, J.~S.}, \au{Blake, T.~D.}, \au{Lockerby, D.~A.} \&
  \au{Sprittles, J.~E.}} \yr{2021}  \at{Putting the micro into the macro: Using
  a molecularly-augmented hydrodynamic model to investigate the flow
  instability of a liquid nano plug}.  \jt{J. Fluid Mech.}  \bvol{To be
  submitted}.

\bibitem[Keeler {\em et~al.\/}(2019)Keeler, Thompson, Lemoult, Juel \&
  Hazel]{keeler2019invariant}
{\sc \au{Keeler, J.~S.}, \au{Thompson, A.~B.}, \au{Lemoult, G.}, \au{Juel, A.}
  \& \au{Hazel, A.~L.}} \yr{2019}  \at{The influence of invariant solutions on
  the transient behaviour of an air bubble in a hele-shaw channel}.  \jt{Proc.
  R. Soc. Lond. A}  \bvol{879},  \pg{1--27}.

\bibitem[Kerswell {\em et~al.\/}(2014)Kerswell, Pringle \&
  Willis]{kerswell2014optimization}
{\sc \au{Kerswell, R.~R.}, \au{Pringle, C. C.~T.} \& \au{Willis, A.~P.}}
  \yr{2014}  \at{An optimization approach for analysing nonlinear stability
  with transition to turbulence in fluids as an exemplar}.  \jt{Rep. Prog.
  Phys.}  \bvol{77}~(8),  \pg{085901}.

\bibitem[Kuznetsov(1998)]{kuznetsov2013elements}
{\sc \au{Kuznetsov, Y.~A.}} \yr{1998} {\em Elements of Applied Bifurcation
  Theory (3rd Ed.)\/}.  \publ{Springer-Verlag}.

\bibitem[Li(2005)]{li2005overview}
{\sc \au{Li, X.~S.x}} \yr{2005}  \at{An overview of superlu: Algorithms,
  implementation, and user interface}.  \jt{ACM Transactions on Mathematical
  Software (TOMS)}  \bvol{31}~(3),  \pg{302--325}.

\bibitem[Liu {\em et~al.\/}(2017)Liu, Carvalho \& Kumar]{liu2017mechanism}
{\sc \au{Liu, C-Y.}, \au{Carvalho, M.~S.} \& \au{Kumar, S.}} \yr{2017}
  \at{Mechanisms of dynamic wetting failure in the presence of soluble
  surfactants}.  \jt{J. Fluid Mech.}  \bvol{825},  \pg{677--703}.

\bibitem[Liu {\em et~al.\/}(2019)Liu, Carvalho \& Kumar]{liu2019predictions}
{\sc \au{Liu, C-Y.}, \au{Carvalho, M.~S.} \& \au{Kumar, S.}} \yr{2019}
  \at{Dynamic wetting failure in curtain coating: Comparison of model
  predictions and experimental observations}.  \jt{Chem. Eng. Sci.}
  \bvol{195},  \pg{74--82}.

\bibitem[Liu {\em et~al.\/}(2016{\natexlab{{\em a\/}}})Liu, Vandre, Carvalho \&
  Kumar]{liu2016assist}
{\sc \au{Liu, C-Y.}, \au{Vandre, E.}, \au{Carvalho, M.~S.} \& \au{Kumar, S.}}
  \yr{2016{\natexlab{{\em a\/}}}}  \at{Dynamic wetting failure and hydrodynamic
  assist in curtain coating}.  \jt{J. Fluid Mech.}  \bvol{808},  \pg{290--315}.

\bibitem[Liu {\em et~al.\/}(2016{\natexlab{{\em b\/}}})Liu, Vandre, Carvalho \&
  Kumar]{liu2016surfactant}
{\sc \au{Liu, C-Y.}, \au{Vandre, E.}, \au{Carvalho, M.~S.} \& \au{Kumar, S.}}
  \yr{2016{\natexlab{{\em b\/}}}}  \at{Dynamic wetting failure in surfactant
  solutions}.  \jt{J. Fluid Mech.}  \bvol{789},  \pg{285--309}.

\bibitem[Oron {\em et~al.\/}(1997)Oron, Davis \& Bankoff]{oron1997long}
{\sc \au{Oron, A.}, \au{Davis, H.~S.} \& \au{Bankoff, G.~S.}} \yr{1997}
  \at{Long-scale evolution of thin liquid films}.  \jt{Rev. Mod. Phys.}
  \bvol{69}~(3),  \pg{931}.

\bibitem[Pack {\em et~al.\/}(2018)Pack, Kaneelil, Kim \& Sun]{pack2018contact}
{\sc \au{Pack, M.}, \au{Kaneelil, P.}, \au{Kim, H.} \& \au{Sun, Y.}} \yr{2018}
  \at{Contact line instability caused by air rim formation under nonsplashing
  droplets}.  \jt{Langmuir}  \bvol{34}~(17),  \pg{4962--4969}.

\bibitem[Reysatt \& Qu{\'e}r{\'e}(2006)]{reysatt2006burst}
{\sc \au{Reysatt, E.} \& \au{Qu{\'e}r{\'e}, D.}} \yr{2006}  \at{Bursting of a
  fluid film in a viscous environment}.  \jt{EPL}  \bvol{73},  \pg{236--247}.

\bibitem[Sackinger {\em et~al.\/}(1996)Sackinger, Schunk \&
  Rao]{sackinger1996newton}
{\sc \au{Sackinger, P.~A.}, \au{Schunk, P.~Randall} \& \au{Rao, R.~R.}}
  \yr{1996}  \at{A newton--raphson pseudo-solid domain mapping technique for
  free and moving boundary problems: a finite element implementation}.  \jt{J.
  Comp. Phys.}  \bvol{125}~(1),  \pg{83--103}.

\bibitem[Sbragaglia {\em et~al.\/}(2008)Sbragaglia, Sugiyama \&
  Biferale]{sbragaglia2008wetting}
{\sc \au{Sbragaglia, M.}, \au{Sugiyama, K.} \& \au{Biferale, L.}} \yr{2008}
  \at{Wetting failure and contact line dynamics in a couette flow}.  \jt{J.
  Fluid Mech.}  \bvol{614},  \pg{471--493}.

\bibitem[Semenov {\em et~al.\/}(2011)Semenov, Starov, Velarde \&
  Rubio]{velarde2011discussion}
{\sc \au{Semenov, S.}, \au{Starov, V.~M.}, \au{Velarde, M.~G.} \& \au{Rubio,
  R.~G.}} \yr{2011}  \at{Droplets evaporation: Problems and solutions}.
  \jt{Euro. Phys. J. Spec. Top.}  \bvol{197}~(1),  \pg{265--278}.

\bibitem[Severtson \& Aidun(1996)]{severtson1996stability}
{\sc \au{Severtson, Y.~C.} \& \au{Aidun, C.~K/}} \yr{1996}  \at{Stability of
  two-layer stratified flow in inclined channels: applications to air
  entrainment in coating systems}.  \jt{J. Fluid Mech.}  \bvol{312},
  \pg{173--200}.

\bibitem[Shikhmurzaev(2007)]{shikhmurzaev2007capillary}
{\sc \au{Shikhmurzaev, Y.~D.}} \yr{2007} {\em Capillary flows with forming
  interfaces\/}.  \publ{CRC Press}.

\bibitem[Snoeijer \& Andreotti(2013)]{snoeijer2013moving}
{\sc \au{Snoeijer, J.~H.} \& \au{Andreotti, B.}} \yr{2013}  \at{Moving contact
  lines: scales, regimes, and dynamical transitions}.  \jt{Ann. Rev. Fluid.
  Mech.}  \bvol{45},  \pg{269--292}.

\bibitem[Snoeijer {\em et~al.\/}(2007)Snoeijer, Andreotti, Delon \&
  Fermigier]{snoeijer2007part1}
{\sc \au{Snoeijer, J.~H.}, \au{Andreotti, B.}, \au{Delon, G.} \& \au{Fermigier,
  M.}} \yr{2007}  \at{Relaxation of a dewetting contact line. part 1. a
  full-scale hydrodynamic calculation}.  \jt{J. Fluid. Mech.}  \bvol{579},
  \pg{63--83}.

\bibitem[Snoeijer {\em et~al.\/}(2006)Snoeijer, Delon, Andreotti \&
  Fermigier]{snoeijer2006avoid}
{\sc \au{Snoeijer, J.~H.}, \au{Delon, G.}, \au{Andreotti, B.} \& \au{Fermigier,
  M.}} \yr{2006}  \at{Avoided critical behavior in dynamically forced wetting}.
   \jt{Phys. Rev. Letters}  \bvol{96}~(174504).

\bibitem[Snoeijer {\em et~al.\/}(2008)Snoeijer, Ziegler, Andreotti, Fermigier
  \& Eggers]{snoeijer2008thick}
{\sc \au{Snoeijer, J.~H.}, \au{Ziegler, J.}, \au{Andreotti, B.}, \au{Fermigier,
  M.} \& \au{Eggers, J.}} \yr{2008}  \at{Thick films of viscous fluid coating a
  plate withdrawn from a liquid reservoir}.  \jt{Phys Rev. Letters}
  \bvol{100}~(24),  \pg{244502}.

\bibitem[Sprittles(2015)]{sprittles2015air}
{\sc \au{Sprittles, J.~E.}} \yr{2015}  \at{Air entrainment in dynamic wetting:
  Knudsen effects and the influence of ambient air pressure}.  \jt{J. Fluid
  Mech.}  \bvol{769},  \pg{444--481}.

\bibitem[Sprittles(2017)]{sprittles2017kinetic}
{\sc \au{Sprittles, J.~E.}} \yr{2017}  \at{Kinetic effects in dynamic wetting}.
   \jt{Phy. Rev. Letters}  \bvol{118}~(11),  \pg{114502}.

\bibitem[Sprittles \& Shikhmurzaev(2011{\natexlab{{\em
  a\/}}})]{sprittles2011viscous2}
{\sc \au{Sprittles, J.~E.} \& \au{Shikhmurzaev, Y.~D.}} \yr{2011{\natexlab{{\em
  a\/}}}}  \at{Viscous flow in domains with corners: Numerical artifacts, their
  origin and removal}.  \jt{Comput. Methods Appl. Mech. Eng.}
  \bvol{200}~(9-12),  \pg{1087--1099}.

\bibitem[Sprittles \& Shikhmurzaev(2011{\natexlab{{\em
  b\/}}})]{sprittles2011viscous1}
{\sc \au{Sprittles, J.~E.} \& \au{Shikhmurzaev, Y.~D.}} \yr{2011{\natexlab{{\em
  b\/}}}}  \at{Viscous flows in corner regions: Singularities and hidden
  eigensolutions}.  \jt{Int. J. Numer. Methods. Fluids}  \bvol{65}~(4),
  \pg{372--382}.

\bibitem[Sprittles \& Shikhmurzaev(2013)]{sprittles2013finite}
{\sc \au{Sprittles, J.~E.} \& \au{Shikhmurzaev, Y.~D.}} \yr{2013}  \at{Finite
  element simulation of dynamic wetting flows as an interface formation
  process}.  \jt{J. Comp. Phys.}  \bvol{233},  \pg{34--65}.

\bibitem[Stay \& Barocas(2003)]{stay2003coupled}
{\sc \au{Stay, M.~S.} \& \au{Barocas, V.~H.}} \yr{2003}  \at{Coupled
  lubrication and stokes flow finite elements}.  \jt{Int. J. Numer. Methods.
  Fluids}  \bvol{42},  \pg{129--146}.

\bibitem[Thoroddsen {\em et~al.\/}(2012)Thoroddsen, Thoraval, Takehara \&
  Etoh]{thoroddsen2012micro}
{\sc \au{Thoroddsen, S.~T.}, \au{Thoraval, M-J.}, \au{Takehara, K.} \&
  \au{Etoh, T.~G.}} \yr{2012}  \at{Micro-bubble morphologies following drop
  impacts onto a pool surface}.  \jt{J. Fluid Mech.}  \bvol{708},
  \pg{469--479}.

\bibitem[Vandre(2013)]{vandrethesis}
{\sc \au{Vandre, E.}} \yr{2013}  \at{Onset of dynamic wetting failure: The
  mechanics of high-speed fluid displacement}. PhD thesis, University of
  Minnesota, USA.

\bibitem[Vandre {\em et~al.\/}(2012)Vandre, Carvalho \& Kumar]{vandre2012delay}
{\sc \au{Vandre, E.}, \au{Carvalho, M.~S.} \& \au{Kumar, S.}} \yr{2012}
  \at{Delaying the onset of dynamic wetting failure through meniscus
  confinement}.  \jt{J. Fluid Mech.}  \bvol{707},  \pg{496--520}.

\bibitem[Vandre {\em et~al.\/}(2013)Vandre, Carvalho \&
  Kumar]{vandre2013mechanism}
{\sc \au{Vandre, E.}, \au{Carvalho, M.~S.} \& \au{Kumar, S.}} \yr{2013}  \at{On
  the mechanism of wetting failure during fluid displacement along a moving
  substrate}.  \jt{Phys. Fluids}  \bvol{25}.

\bibitem[Voinov(1976)]{voinov1976}
{\sc \au{Voinov, O.~V.}} \yr{1976}  \at{Hydrodynamics of wetting}.  \jt{Fluid
  Dyn.}  \bvol{11},  \pg{714–721}.

\bibitem[Zienkiewicz \& Zhu(1992)]{Zienkiewicz1992}
{\sc \au{Zienkiewicz, O.~C.} \& \au{Zhu, J.~Z.}} \yr{1992}  \at{The
  superconvergent patch recovery and a posteriori error estimates. {P}art 1:
  The recovery technique}.  \jt{Intl J. Numer. Meth. Engng}  \bvol{33}~(7),
  \pg{1331--1364}.

\end{thebibliography}
\createbib{dynamic_wetting1}

\end{document}